\documentclass[aps, superscriptaddress, prb, twocolumn]{revtex4-1}
\usepackage{amsmath}
\usepackage{graphicx}
\usepackage{wasysym}
\usepackage{amsfonts}
\usepackage{amssymb}
\usepackage{amsfonts}
\usepackage{bm}
\usepackage{natbib}
\usepackage{enumerate}
\usepackage{color}
\usepackage{hyperref}

\usepackage{times}
\newcommand{\beq}{\begin{equation}}
\newcommand{\eeq}{\end{equation}}
\newcommand{\bea}{\begin{eqnarray}}
\newcommand{\eea}{\end{eqnarray}}
\newcommand{\bes}{\begin{split}}
\newcommand{\ees}{\end{split}}

\newcommand{\la}{\langle}
\newcommand{\ra}{\rangle}

\mathchardef\nss="711B

\def\nss{\mathcal{S}}

\newcommand{\normord}[1]{:\mathrel{#1}:}

\def\be{\begin{eqnarray}}
\def\ee{\end{eqnarray}}
\def\beq{\begin{equation}}
\def\eeq{\end{equation}}

\newlength{\myL}

\begin{document}

\title{Chiral Sachdev-Ye model: Integrability and chaos of anyons in 1+1d}
\affiliation{Department of Physics, Princeton University, Princeton, New Jersey 08544, USA}
\affiliation{The Rudolf Peierls Centre for Theoretical Physics, University of Oxford, Oxford OX1 3PU, United Kingdom}
\author{Yichen Hu$^{1,2}$}
\author{Biao Lian$^{1}$}

\begin{abstract}
We construct and study a chiral Sachdev-Ye (SY) model consisting of $N$ chiral SU$(M)_1$ Wess-Zumino-Witten (WZW) models with current-current interactions among each other, which generalizes the 0+1d quantum chaotic SY spin model into 1+1d chiral system with anyon excitations. Each WZW model hosts Abelian anyons as charge excitations, and may arise as the chiral edge theory of 2+1d gapped topological phases. We solve the chiral SY model in two limits which show distinct quantum dynamics. The first limit is the case with uniform interactions at any integers $N$ and $M$, which is integrable and decomposes into a chiral SU$(M)_N$ WZW model and its coset with a different ``speed of light". When $N=M=2$, the model maps to a free Majorana fermion model. The second limit is the large $N$ and $M$ limit with random interactions, which is solvable to the leading $\frac{1}{NM}$ order, and exhibits many-body quantum chaos in the out-of-time-ordered correlation of anyons. As the interaction strength approaches the upper limit preserving the chirality, the leading velocity-dependent Lyapunov exponent of the model saturates the maximal chaos bound $2\pi/\beta$ at temperature $\beta^{-1}$. 
\end{abstract}
\maketitle

\section{Introduction}

In the past few years, the quantum chaos of many-body systems has attracted extensive studies, which is believed to be crucial for achieving thermal equilibration. Of particular interest is the 0+1 dimensional (0+1d) Sachdev-Ye-Kitaev (SYK) model \cite{SY93,Polchinski:2016xgd,Maldacena:2016hyu,Kitaev:2017awl,Gu2020} and its generalizations \cite{Witten:2016iux,gurau,Klebanov:2016xxf,Gu:2016oyy,Berkooz:2016cvq,davison2017,Jian:2017unn,chen2017,cai2018,zhangp2018,Turiaci:2017zwd,Blake:2017,Song2017,Das2018,Liu2019,Murugan:2017eto,Berkooz:2017efq,Narayan2017,Giombi:2017dtl,KPT,Gross2017,Banerjee2017,kim2019,Klebanov2020,Ahn2019,Peng2017}, which are analytically solvable by large $N$ expansion techniques and are many-body quantum chaotic. The SYK-type physics has its root in condensed matter physics. The very first analytically solvable model is the pioneer work proposed by Sachdev and Ye \cite{SY93} in the form of a 0+1d spin model of $N$ randomly interacting SU($M$) spins, known as the Sachdev-Ye (SY) model:
\begin{equation}\label{eq:0dSY}
H_{\text{SY}}=\sum_{i\neq j}^N J_{ij}\mathbf{S}_i\cdot\mathbf{S}_j\ ,
\end{equation}
where $\mathbf{S}_i$ are SU($M$) spins, and $J_{ij}$ are random interactions. In the large $N$ and $M$ limit, the SY model maps to a fermion model which behaves the same as the $q=4$ SYK model in the large $N$ limit. The quantum chaos of the SYK-type models are characterized by the quantum Lyapunov exponent of the out-of-time ordered correlation (OTOC). At low temperatures $T=\beta^{-1}$, the SYK model and the SY model in 0+1d exhibit a maximal Lyapunov exponent $\lambda=2\pi/\beta$, reaching the upper bound for quantum chaotic system \cite{Shenker2014,Shenker2015,Maldacena:2015waa}. Moreover, the 0+1d SYK model has a holographic dual to a quantum gravity theory in 1+1d spacetime, which has generated enormous interest of study \cite{Turiaci:2017zwd,Sachdev:2010prl,Almheiri:2014cka,Jensen:2016pah,Maldacena:2017axo,Engelsoy:2016xyb,Maldacena:2016upp,Maldacena:2018lmt,kim2019,Cotler:2016fpe,Saad:2018bqo,Saad:2019lba,Sachdev:2010prl}.

The generalization of the SYK model has been going in several different directions. The first direction is generalizations into fermion models in higher dimensions. Unlike in 0+1d where the interactions are relevant, the fermion interactions in 1+1d or higher dimensions are usually irrelevant, in which case the SYK-type physics will be absent. For instance, the random four-fermion interactions of the 1+1d nonchiral fermions are shown to be marginally irrelevant in the large $N$ limit \cite{Berkooz:2017efq}. Therefore, to achieve relevant interactions and realize the SYK-type physics, unrealistic or fine-tuned kinetic term for the fermions is usually assumed \cite{Gu:2016oyy,Berkooz:2016cvq,davison2017,Jian:2017unn,chen2017,cai2018,zhangp2018,Turiaci:2017zwd,Blake:2017,Song2017,Das2018}. There is, however, an intriguing exception, which is the chiral fermions in 1+1d with the usual linear kinetic Hamiltonian, where the local chiral four-fermion interactions are fixed to have an exactly marginal scaling dimension. It is shown in Ref. \cite{Lian2019} that the generalization of SYK model into 1+1d chiral Majorana fermions with the marginal random four-fermion interactions, coined ``the chiral SYK model'', exhibits the SYK-type many-body quantum chaos at all energy scales. The second direction of generalization is the class of tensor models which contains no randomness but shares the same large $N$ physics with the SYK model \cite{Narayan2017,Klebanov:2016xxf,Giombi:2017dtl,Witten:2016iux,gurau,KPT}. This yields a generic way to write a non-random equivalence to a model exhibiting SYK-type physics. The third direction of generalization is to models of particles other than fermions. For instance, bosonic SYK models with $q$-boson interactions have been studied in 1+1d \cite{Murugan:2017eto,Liu2019}, for which interactions are relevant. However, generalizations of SYK-type models into anyonic systems have not been systematically explored yet.

This paper is aimed at forging the first and third directions above to find a higher dimensional anyonic generalization of the 0+1d SYK models. For this purpose, 1+1d chiral systems are natural candidates, as they allow chiral interactions which can lead to the SYK-type quantum chaos, and can host anyons as excitations. In reality, 1+1d chiral systems generically describe the topologically protected chiral edge states of 2+1d gapped topological phases of matter such as fractional quantum Hall (FQH) states. The interaction effects and quantum chaos of such 1+1d chiral systems are of importance in understanding the edge transport phenomena in experiments. In literature, these chiral edge states are often described by chiral Luttinger liquids and other conformal field theories (CFTs), which are integrable models and do not show quantum chaos. Therefore, the chiral edge states are often believed to be fairly coherent and can exhibit interference among each other in edge interferometry devices. For instance, the Fabry-P\'erot edge state interferometer \cite{Bartolomei2020,Carrega2021,McClure2012,Ofek2010,Halperin2011} has recently been employed to detect the anyon braiding phase in FQH systems. Understanding the conditions for quantum integrability and quantum chaos in generic 1+1d chiral models is thus significant for interferometer experiments studying topological edge states.

As a first step towards this goal, in this paper we propose and study a 1+1d model we call the \emph{chiral SY model}, which is a minimal generalization of the 0+1d SY model (Eq. (\ref{eq:0dSY})) into the chiral anyonic systems. The key idea is to generalize the $N$ mutually interacting 0+1d SU($M$) spins of the original SY model into $N$ 1+1d chiral SU$(M)_1$ WZW models with marginal chiral current-current interactions among each other. Such WZW models may arise as the edge theories of 2+1d gapped topological phases (e.g., multilayer bosonic FQH states). In particular, the marginal chiral current-current interactions have conformal spin two,  and thus break the conformal symmetry of the WZW models, although the scaling symmetry is still preserved \cite{Lian2019}. As a result, there is no unique ``speed of light" in the system. The minimal charge excitation in each WZW model has a conformal spin $\Delta=\frac{M-1}{2M}$ and an anyonic statistical phase $2\pi\Delta$. We study two analytically solvable limits of the chiral SY model, which are in the quantum integrable and quantum chaotic regimes, respectively. The first limit is when the interactions $J_{ij}$ between different WZW models are uniform, in which case we will show that the chiral SY model is integrable: it decomposes into a sum of a chiral SU$(M)_N$ WZW model and its coset with different speeds of light.  In the minimal case with $N=M=2$, free chiral Majorana fermions with different Fermi velocities emerge as quasiparticles, which can be understood as the composite particles of two semions (Abelian anyons at $M=2$ with minimal charge excitation). The second limit is the large $N$ and $M$ limit with random interactions $J_{ij}$, which is solvable to the leading $\frac{1}{NM}$ order. We will show that the OTOC of anyons in this case grows exponentially at all energy scales, exhibiting many-body quantum chaos. As the interaction strength approaches the upper limit preserving the edge chirality, the Lyapunov exponent of the OTOC at temperature $\beta^{-1}$ approaches the maximal chaos bound \cite{Shenker2014,Shenker2015,Maldacena:2015waa} $2\pi/\beta$. Intriguingly, by introducing a set of ancillary fields and enlarging the Hilbert space, the chiral SY model can be mapped to an interacting chiral fermion model which resembles the complex version of the chiral SYK model studied in Ref. \cite{Lian2019}. As a result, all the correlations (e.g., the OTOC) of anyons factorizes into certain correlations of the interacting chiral fermions and a decoupled free vertex operator correlation governing the anyonic statistical phases, which greatly simplifies their calculations.

The rest of the paper is organized as follows. In Sec. \ref{sec:themodel}, we introduce the 1+1d chiral SY model consisting of $N$ coupled SU$(M)_1$ WZW models written in the bosonized form. Sec. \ref{sec:uniformcoupling} is devoted to solving the chiral SY model with uniform interactions and showing its integrability. The $N=M=2$ case is analyzed in details to show how anyon correlations can be calculated. In Sec. \ref{sec:randomcoupling}, we study the chiral SY model in the large $N$ and $M$ limit with random interactions, and show the OTOC of anyons in this case gives a positive velocity-dependent Lyapunov exponent (VDLE) in the entire causality cone. Lastly, in Sec. \ref{sec:discussion} we summarize the generic methods used in our model for deriving the anyon correlations, and discuss the extension to other chiral anyonic systems.

\section{The model}\label{sec:themodel}

The straightforward chiral 1+1d generalization of the 0+1d SY model is $N$ copies of the SU($M$) level one (denoted as SU$(M)_1$ hereafter) chiral Wess-Zumino-Witten (WZW) model, where $M\ge2$, $N\ge1$ are integers, and the SU$(M)_1$ chiral WZW models can be viewed as the generalization of the SU$(M)$ spins in the $0+1$d SY model. The SU$(M)_1$ chiral WZW model is known to characterize the edge theory of a SU($M$) chiral spin liquid \cite{Tu2014,kalmeyer1987,kalmeyer1989,wen1989}, or a filling $\nu=\frac{M-1}{M}$ bosonic Halperin FQH state with $M-1$ boson flavors \cite{Blok1992} (see Appendix.~\ref{A}). In particular, the SU(2)$_1$ and SU(3)$_1$ WZW models correspond to the edge theories of the $m=2$ bosonic Laughlin state and the $(2,2,1)$ bosonic Halperin state, respectively. Accordingly, the model hosts a set of Abelian anyons.

We construct our model by considering $N$ identical copies of the chiral SU$(M)_1$ WZW model, and adding SU($M$) symmetric interactions among them. Focusing on the low energy physics, we ignore all irrelevant terms, which restricts the generic local Lagrangian density of the system into two parts:
\begin{equation}\label{SY}
\mathcal{L}(t,x)=\mathcal{L}_0(t,x)+\mathcal{L}_{\text{int}}(t,x)\ ,
\end{equation}
where $x$ and $t$ are the spatial and time coordinates, respectively. The first term $\mathcal{L}_0$ gives the SU$(M)_1$ chiral WZW Lagrangian within each copy, which can be written as a chiral Luttinger liquid Lagrangian of $N(M-1)$ free chiral boson fields $\phi_{i\mu}$ ($1\le i\le N$, $1\le \mu\le M-1$) as:
\begin{equation}\label{eq:L0}
\mathcal{L}_0=-\frac{1}{4\pi} \sum_{i=1}^N\sum_{\mu,\nu=1}^{M-1}K_{\text{SU}(M)_1}^{\mu\nu}\partial_x \phi_{i\nu}( \partial_t +\partial_x) \phi_{i\mu}\ ,
\end{equation}
with $\phi_{i\mu}$ identified with $\phi_{i\mu}+2\pi$. Here $K_{\text{SU}(M)_1}$ is the $K$-matrix of the WZW model, which is given by the Cartan matrix of the SU($M$) group $K_{\text{SU}(M)_1}^{\mu\nu}=2\delta_{\mu,\nu}-\delta_{\mu,\nu-1}-\delta_{\mu,\nu+1}$ for $1\le \mu,\nu\le M-1$. Accordingly, the chiral boson fields satisfy the equal-time commutation relation
\begin{equation}
[\phi_{i\mu}(x), \phi_{j\nu}(x')]=i\pi \text{sgn}(x-x') \left(K_{\text{SU}(M)_1}^{-1}\right)_{\mu\nu}\delta_{ij}\ ,
\end{equation}
where $\text{sgn}(x)$ is the sign function. In $\mathcal{L}_0$, all the chiral boson fields have the same velocity (``speed of light") which we have normalized to $1$, as constrained by the SU($M$) symmetry within each copy $i$ and our assumption that the $N$ copies of WZW models are identical.

The second term $\mathcal{L}_{\text{int}}$ is an SU$(M)$ invariant chiral current-current interaction between different copies:
\begin{equation}\label{eq:Lint}
\mathcal{L}_{\text{int}}=-\sum_{i\neq j}^N J_{ij}\bm{\mathcal{S}}_i(t,x) \cdot \bm{\mathcal{S}}_j(t,x)\ ,
\end{equation}
where $\bm{\mathcal{S}}_i(t,x)$ is the chiral SU$(M)_1$ current density of the $i$-th copy in the vector form, which has $M^2-1$ components $\mathcal{S}^a_i(t,x)$ ($1\le a\le M^2-1$) corresponding to the $M^2-1$ generators of the SU($M$) group. The interaction constants $J_{ij}=J_{ji}$ are real and spatially uniform, and symmetric in the copy indices $i,j$. With the level being one, the current density satisfies the equal-time Kac-Moody algebra \cite{DiFrancesco1997},
\beq\label{eq:KacMoody}
[\mathcal{S}^a_i(x),\mathcal{S}^b_i(x')]=if^{abc}\mathcal{S}^c_i(x) \delta(x-x')+\frac{i\delta^{ab}}{4\pi}\partial_x \delta(x-x')
\eeq 
on a spatial slice at any given time $t$ (the time variable $t$ is omitted for simplicity), where $f^{abc}$ is the structure constant of the $SU(M)$ group. The current density $\bm{\mathcal{S}}_i(t,x)$ can also be expressed in terms of the boson fields $\phi_{i\mu}$. It is clear that the chiral model in Eq. (\ref{SY}) has a global SU$(M)$ symmetry.

We see that the chiral model in Eq. (\ref{SY}) can be viewed as a natural minimal 1+1d generalization of the 0+1d SY spin model \cite{SY93} with $N$ interacting SU($M$) spins with spin-spin interactions $J_{ij}$, where the SU$(M)_1$ current density $\bm{\mathcal{S}}_i(t,x)$ is the 1+1d generalization of the spin operator of a 0+1d spin in the fundamental SU($M$) representation. However, there are two significant differences between our chiral 1+1d model and the 0+1d SY spin model: 

(i) The chiral current-current interaction in 1+1d given by Eq. (\ref{eq:Lint}) has a marginal scaling dimension $2$, since the current density $\bm{\mathcal{S}}_i(t,x)$ has scaling dimension $1$, as can be verified from Eq. (\ref{eq:KacMoody}). As chiral operators, their scaling dimension is strictly fixed and does not have flow under renormalization group analysis \cite{Lian2019}. Therefore, we expect the physics of our model to be energy-scale independent. In contrast, the spin-spin interaction in the 0+1d SY spin model is relevant, which is known to dominate the low-energy physics.

(ii) Charge excitations of the $i$-th copy of our chiral 1+1d model here are $M-1$ Abelian anyons $\{\chi_{i,n}\}_{n=1}^{M-1}$ with conformal spin $\Delta_{i,n}=\frac{n(M-n)}{2M}$ and statistical angle $2\pi \Delta_{i,n}$. The fusion structure of these anyons are straightforward: $\chi_{i,n} \times \chi_{i,n'}= \chi_{i,(n+n') \mod M}. $ This allows us to study the quantum chaos of anyons, which has not been studied before. This is different from the 0+1d SY spin model, where there are no anyonic excitations.

We also note that the nonzero conformal spin of the interaction in Eq. (\ref{eq:Lint}) breaks the conformal symmetry of the original WZW models in Eq. (\ref{eq:L0}), although the scaling symmetry is still preserved. As a result, there is no unique ``speed of light" in the model.

As we will show below, the 1+1d chiral model in Eq. (\ref{SY}) is analytically solvable in two different limits. The first limit is the uniform interaction limit for any $N$ and $M$, where $J_{ij}$ independent of $i,j$, in which case the model is integrable. The second limit is the large $N$ and $M$ limit with random interactions $J_{ij}$, which exhibits quantum chaos analogous to that of the 1+1d chiral SYK model \cite{Lian2019}, but is for anyonic excitations.

\section{Uniform interaction: an integrable point}\label{sec:uniformcoupling}

In this section, we focus on the case with uniform interactions:
\begin{equation}
J_{ij}=J\ .
\end{equation}
We will show that the model with uniform interactions is integrable. Particularly, we will study the simplest $N=2,M=2$ case in detail,  to understand the integrable structure and the anyon correlations in this case.

\subsection{Exact solution for $N=2, M=2$}\label{sec:N=2M=2freeMajorana-0}
At $N=2, M=2$, the chiral model of Eq. (\ref{SY}) consists of two copies of SU$(2)_1$ chiral WZW model, and only has one interaction interaction $J_{12}=J$. Such a model may describe the edge theory of two filling $\nu=1/2$ bosonic Laughlin FQH states with an edge interaction $J$, and hosts semions which have an anyonic statistical phase $\pi/2$. By Eq. (\ref{eq:L0}), the $N=2, M=2$ model Lagrangian has a one-by-one $K$-matrix $K_{\text{SU}(2)_1}=2$, and contains only two bosonic fields $\phi_i$ ($i=1,2$) satisfying the commutation relation 
\begin{equation}\label{eq:comm-N=2M=2}
[\phi_{i}(x), \phi_{j}(x')]=i\frac{\pi}{2} \text{sgn}(x-x')\delta_{ij}\ .
\end{equation}
The three components of the SU$(2)_1$ current operator $\bm{\mathcal{S}}_i(x)$ of the $i$-th copy can then be written as $\mathcal{S}_i^{\pm}=\mathcal{S}_i^x\pm i\mathcal{S}_i^y=e^{\mp 2i \phi_i}$, and $\mathcal{S}_i^z=\frac{\partial_x \phi_i}{2\pi}$.

To be more generic, here we will relax the SU(2) symmetry of the model, assuming an anisotropic interaction between the two copies $\mathcal{L}_\text{int}=-2J_\perp (\mathcal{S}_1^x\mathcal{S}_2^x+\mathcal{S}_1^y\mathcal{S}_2^y)-2J_z\mathcal{S}_1^z\mathcal{S}_2^z$. The SU(2) invariant interaction in Eq. (\ref{eq:Lint}) is recovered when $J_\perp=J_z=J$. The total Lagrangian density $\mathcal{L}=\mathcal{L}_0+\mathcal{L}_\text{int}$ is therefore given by
\begin{equation}\label{eq:LN=2M=2}
\begin{split}
\mathcal{L}=&-\frac{2}{4\pi} \sum_{i=1}^2\left[\partial_t \phi_i \partial_x \phi_i+(\partial_x \phi_i)^2 \right]\\
&-\frac{J_z}{2\pi^2}\partial_x \phi_1\partial_x \phi_2 
-J_\perp \left(e^{2i\phi_1-2i\phi_2}+H.c.\right)\ .
\end{split}
\end{equation}

\subsubsection{Mapping to free Majorana fermions}\label{sec:N=2M=2freeMajorana}

Despite having interactions between boson fields, the model in Eq. (\ref{eq:LN=2M=2}) can be mapped to a model of free Majorana fermions by refermionization. We first define the following boson field basis transformation and complex fermion annihilation operators $c_{\pm}(t,x)$ as
\beq\label{eq:N=2M=2fermion-def}
\phi_{\pm}=\phi_1 \pm \phi_2\ ,\qquad c_{\pm}=e^{i\phi_{\pm}}\ ,
\eeq 
where the vertex operators $e^{i\phi_{\pm}}$ have the correct scaling dimension $1/2$ of fermion operators. Using the mappings of fermion bilinear $-ic_\eta^\dag \partial_x c_\eta=\frac{1}{4\pi}(\partial_x \phi_\eta)^2$ and $-ic_\eta\partial_x c_\eta=2\pi e^{2i\phi_\eta}$ for $\eta=\pm$ (see Appendix.~\ref{app:B}), we can rewrite the Lagrangian density in Eq. (\ref{eq:LN=2M=2}) as
\begin{equation}
\mathcal{L}=\sum_{\eta=\pm}ic_\eta^\dagger(\partial_t+u_\eta \partial_x)c_\eta +\frac{J_\perp}{2\pi} (ic_-^\dagger\partial_x c_-^\dagger+H.c.)\ ,
\end{equation}
where we have defined velocities $u_\pm=1\pm\frac{J_z}{2\pi}$. Therefore, we see that the model maps to a free fermion model with a superconducting pairing term proportional to $J_\perp$, which is clearly integrable. 

By redefining $c_-=\frac{1}{\sqrt{2}}(\gamma_{1,-}+i\gamma_{2,-})$ with two Majorana fermion fields $\gamma_{1,-}$ and $\gamma_{2,-}$, the Lagrangian density becomes
\begin{equation}\label{eq:LfreeMajorana}
\mathcal{L}=ic_+^\dagger(\partial_t+u_+ \partial_x)c_+ +\sum_{j=1,2}\frac{i}{2}\gamma_{j,-}(\partial_t+u_{j,-} \partial_x)\gamma_{j,-}\ ,
\end{equation}
where the velocities are given by 
\begin{equation}\label{eq:N=M=2-Majarana-velocity}
u_+=1+\frac{J_z}{2\pi}, \quad u_{j,-}=1+(-1)^{j-1}\frac{J_\perp}{\pi}-\frac{J_z}{2\pi}.
\end{equation}
Therefore, we see that the $N=2, M=2$ interacting chiral model reconstructs itself into a free chiral complex fermion mode and two free chiral Majarana modes. The zero-temperature two-point functions of the fermion fields are then given by
\begin{equation}\label{eq:free-fermion-corre1}
\langle c_+(t,x)c_+^\dag(0,0)\rangle =\frac{1}{2\pi i}\frac{1}{(u_+t-x-i0^+)}\ ,
\end{equation}
and
\begin{equation}\label{eq:free-fermion-corre2}
\langle \gamma_{j,-}(t,x)\gamma_{j,-}(0,0)\rangle =\frac{1}{2\pi i}\frac{1}{(u_{j,-}t-x-i0^+)}\ ,
\end{equation}
where $0^+$ stands for a positive infinitesimal number.

\begin{figure}[tbp]
\centering\includegraphics[width=0.5\textwidth]{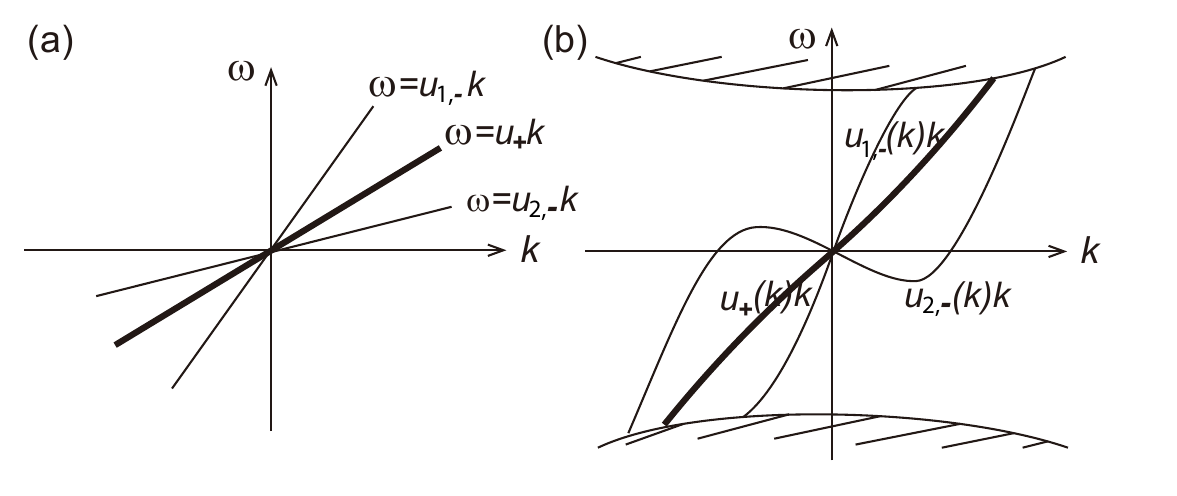}
\caption{(a) Illustration of dispersions $\omega=u_+k$ of the chiral complex fermion mode $c_+$ and $\omega=u_{1,-} k, u_{2,-} k$ of the chiral Majorana modes $\gamma_{1,-}, \gamma_{1,-}$ in Eq. (\ref{eq:LfreeMajorana}). (b)  If $\frac{|J_\perp|}{\pi}+\frac{J_z}{2\pi}>1$ or $\frac{J_z}{2\pi}<-1$, some of the three velocities in (a) will become negative, and the chirality of the system is not preserved. If all chiral modes live at the edge of a gapped bulk topological condensed matter system (the shaded area in (b) represents bulk states), the net chiral central charge (number of chiral modes) must be preserved. Thus, edge reconstruction would take place in the UV, bending the negative velocity modes back to positive velocities at large momentum $k$. (For both (a) and (b), we plot the case with $J_\perp>J_z$.)}\label{fig:largeK}
\end{figure}

When the SU(2) symmetry is recovered, namely, $J_\perp=J_z=J$, we see that the two velocities $u_+=u_{j,-}=1+\frac{J}{2\pi}$ become equal. Therefore, the Lagrangian density in Eq. (\ref{eq:LfreeMajorana}) can be viewed as consisting of three identical chiral Majorana fermion modes with velocity $u_+=1+\frac{J}{2\pi}$ and one chiral Majorana fermion mode with velocity $u_{2,-}=1-\frac{3J}{2\pi}$. Such a model can be understood as a chiral SU(2)$_2$ WZW model with speed of light $u_+$ and an chiral Ising model with speed of light $u_{2,-}$. A more general discussion from this perspective will be given in Sec. \ref{sec:uniform-arbitraryNM}.

We also note that the velocities in Eq. (\ref{eq:LfreeMajorana}) may become negative when $J_\perp$ or $J_z$ is large, which seems to violate the conservation of chirality (number of chiral modes) if the model describes the edge of a gapped topological phase. In this case, as discussed in Ref.  \cite{Lian2019}, the irrelevant terms are nonnegaligible, which will bend the negative slope dispersion back to positive slope at ultraviolet (UV) momenta, as shown in Fig. \ref{fig:largeK}(b), leading to a partially chiral edge model with the same total chirality. In this paper, we shall restrict ourselves to nonnegative velocities, namely, $\frac{|J_\perp|}{\pi}+\frac{J_z}{2\pi}<1$ and $\frac{J_z}{2\pi}>-1$ in this section, so that no UV physics is involved.

\subsubsection{Anyon correlations}

The non-trivial excitations of the $N=2$ copies of chiral SU(2)$_1$ WZW model are anyons known as the semions, the annihilation field operators of which are given by $\chi_i=e^{i\phi_i}$. According to Eq. (\ref{eq:comm-N=2M=2}), the semion fields at equal time satisfy $\chi_i(x)\chi_j(x')=e^{-i\frac{\pi}{2}\delta_{ij}\text{sgn}(x-x')}\chi_i(x')\chi_j(x)$, thus they have a self-statistical phase $\pi/2$. We have seen in the above Sec. \ref{sec:N=2M=2freeMajorana} that the composition of two semions from different copies behaves as free (Majorana) fermions, namely, $\chi_1(x)\chi_2(x)=c_+(x)$ and $\chi_1(x)\chi_2^\dag(x)=c_-(x)$ according to Eq. (\ref{eq:N=2M=2fermion-def}), the correlations of which are given by Eqs. (\ref{eq:free-fermion-corre1}) and (\ref{eq:free-fermion-corre2}).

Now we want to derive the two-point correlations of individual semion fields in this integrable model. The Majorana fermion form of the Lagrangian in Eq. (\ref{eq:LfreeMajorana}) suggests this can be done by adding a duplicate copy of the original model, which yields a total Lagrangian density

\beq\label{eq:Ldup}
\mathcal{L}_{\text{dup}}=\mathcal{L}(\phi_1,\phi_2)+\mathcal{L}(\phi'_1,\phi'_2)\ ,
\eeq 
where $\mathcal{L}(\phi_1,\phi_2)$ is given by Eq. (\ref{eq:LN=2M=2}), and $\phi_j'$ are duplicate boson fields independent of $\phi_j$. 
Noting that $e^{\pm i\phi_j\pm i\phi_j'}$ has scaling dimension $1/2$, we can define a set of fermion annihilation operators
\begin{equation}\label{eq:dup-fermion-def}
c_{j\pm}=e^{i\phi_j\pm i\phi_j'}\ , \qquad (j=1,2).
\end{equation}
Note that the fermions $c_{j\pm}$ here are different from the fermions $c_\pm$ defined in Eq. (\ref{eq:N=2M=2fermion-def}). Since the Lagrangian of $\phi_j$ and $\phi_j'$ are decoupled and identical, we can rewrite the anyon two-point function as 
\begin{equation}\label{eq:semion-2p-dup}
\begin{split}
&\langle \chi_i  (t,x)\chi^\dagger_j(0,0)\rangle=\langle e^{i\phi_i (t,x)}e^{-i\phi_j (0,0)}\rangle \\
=&\langle e^{i\phi_i (t,x)+i\phi_i'(t,x)}e^{-i\phi_j (0,0)-i\phi_j' (0,0)}\rangle^{1/2}\\
=&\langle c_{i+}(t,x)c^\dagger_{j+}(0,0)\rangle^{1/2}\ .
\end{split}
\end{equation}
Therefore, the problem reduces to the calculation of the two-point function of the fermion fields $c_{j,+}$ and $c_{j,+}^\dag$.

From Eq. (\ref{eq:dup-fermion-def}), we can refermionize the duplicated Lagrangian density in Eq. (\ref{eq:Ldup}) into an interacting fermion model:
\begin{equation}\label{eq:Ldup-c}
\begin{split}
\mathcal{L}_{\text{dup}}&=\sum_{j=1,2,\eta=\pm} ic^\dag_{j\eta}(\partial_t+\partial_x)c_{j\eta} \\
&-\frac{J_\perp}{2} (c_{1+}c_{1-}c_{2+}^\dagger c_{2-}^\dagger+c_{1+}c_{1-}^\dagger c_{2+}^\dagger c_{2-}+H.c.) \\
&-\frac{J_z}{2}(c^\dagger_{1+}c_{1+}c^\dagger_{2+}c_{2+}+c^\dagger_{1-}c_{1-}c^\dagger_{2-}c_{2-})\ .
\end{split}
\end{equation}
To solve the duplicated model, we first change to a new fermion basis $d_{\pm,\rho}=\frac{1}{\sqrt{2}}(c_{1\pm}+ic_{2\pm})$ and $d_{\pm,\sigma}=\frac{1}{\sqrt{2}}(c_{1\pm}-ic_{2\pm})$. One can show that the interaction terms become solely bilinear functions of the fermion density operators $d_{\eta,\rho}^\dag d_{\eta,\rho}$ and $d_{\eta,\sigma}^\dag d_{\eta,\sigma}$ ($\eta=\pm$), and the kinetic terms remain diagonal in each fermion species $d_{\pm,\rho/\sigma}$. Therefore, the fermion number of each species $d_{\pm,\rho/\sigma}$ is conserved. We then bosonize the model by defining $d_{\pm,\rho}=e^{i\theta_{\pm,\rho}}$ and $d_{\pm,\sigma}=e^{i\theta_{\pm,\sigma}}$, which imply the mapping $d_{\eta,\alpha}^\dag d_{\eta,\alpha}=\frac{1}{2\pi}\partial_x\phi_{\eta,\alpha}$ ($\eta=\pm,\alpha=\rho,\sigma$), so the Lagrangian density $\mathcal{L}_{\text{dup}}$ in Eq. (\ref{eq:Ldup-c}) becomes a bilinear function of boson fields $\partial_x\theta_{\pm,\rho}$ and $\partial_x\theta_{\pm,\sigma}$ (see Appendix.~\ref{app:C1} for detailed derivations). This converts the model into a free boson model. Lastly, by changing to a new boson basis
\begin{equation}\label{triality}
\widetilde{\theta}_{\eta\eta'}=(\theta_{+,\rho}+\eta\theta_{+,\sigma}+\eta'\theta_{-,\rho}+\eta\eta'\theta_{-,\sigma})/2\ ,
\end{equation}
where $\eta,\eta'=\pm$, we can diagonalize the free boson Lagrangian density into

\beq\label{eq:Ldup-diag}
\mathcal{L}_{\text{dup}}=-\frac{1}{4\pi} \sum_{\eta,\eta'=\pm} \partial_x \widetilde{\theta}_{\eta\eta'} (\partial_t+u_{\eta\eta'} \partial_x) \widetilde{\theta}_{\eta\eta'}\ ,
\eeq 
where the velocities are given by $u_{++}=u_{+-}=1+\frac{J_z}{2\pi}$, and $u_{-\pm}=1\pm\frac{J_\perp}{\pi}-\frac{J_z}{2\pi}$. Note that the four velocities here are related to the three velocities in Eq. (\ref{eq:LfreeMajorana}) by $u_{++}=u_{+-}=u_+$, $u_{-+}=u_{1,-}$ and $u_{--}=u_{2,-}$. If we further define fermion annihilation fields $c_{\eta\eta'}=e^{i\widetilde{\theta}_{\eta\eta'}}$, we see that the Lagrangian density of Eq. (\ref{eq:Ldup-diag}) is fermionized into four free complex fermion modes $f_{\eta\eta'}$ with velocities $u_{\eta\eta'}$, which is exactly two copies of the Lagrangian density before duplication in Eq. (\ref{eq:LfreeMajorana}), as it should be. We also note that the transformation from the fermion basis $d_{\pm,\rho/\sigma}=e^{i\theta_{\pm,\rho/\sigma}}$ to the fermion basis $c_{\eta\eta'}=e^{i\widetilde{\theta}_{\eta\eta'}}$ through Eq. (\ref{triality}) is exactly the triality transformation among the chiral spinor and vector representations of the SO(8) group, which has been employed in various physical contents \cite{shankar1981,shankar1983,Kitaev2010,Maldacena1997,Ryu2012,Kane2020}.

The free boson action in Eq. (\ref{eq:Ldup-diag}) yields a zero-temperature boson two-point function
\begin{equation}
\langle \widetilde{\theta}_{\eta\eta'}(t,x)\widetilde{\theta}_{\zeta\zeta'}(0,0)\rangle =-\delta_{\eta\zeta}\delta_{\eta'\zeta'}\ln [2\pi i(u_{\eta\eta'}t-x-i0^+)]
\end{equation}
up to constant shifts of the boson fields $\widetilde{\theta}_{\eta\eta'}$, which allows us to derive the semion two-point function defined in Eq. (\ref{eq:semion-2p-dup}). First, by noting that the fermion basis $d_{\pm,\rho/\sigma}=e^{i\frac{\pm\widetilde{\theta}_{++}\pm\widetilde{\theta}_{+-}\pm\widetilde{\theta}_{-+}\pm\widetilde{\theta}_{--}}{2}}$, we find the fermion correlation from the Wick's theorem for bosons as
\begin{equation}
\begin{split}
&\langle d_{\zeta,\alpha}(t,x)d_{\zeta,\alpha}^\dag(0,0)\rangle =\prod_{\eta,\eta'=\pm} e^{\frac{1}{4}\langle \widetilde{\theta}_{\eta\eta'}(t,x)\widetilde{\theta}_{\eta\eta'}(0,0)\rangle} \\
=&\frac{1}{2\pi i}\prod_{\eta,\eta'=\pm} (u_{\eta\eta'}t-x-i0^+)^{-1/4}
\end{split}
\end{equation}
for any $\zeta=\pm$ and $\alpha=\rho,\sigma$. In addition, all the correlations between different fermion species $d_{\pm,\rho/\sigma}$ are zero, since the fermion number of each species is conserved. Therefore, the semion two-point function of the original model in Eq. (\ref{eq:LN=2M=2}) is given by
\begin{equation}\label{eq:semion-2p-1}
\begin{split}
&\langle \chi_j  (t,x)\chi^\dagger_k(0,0)\rangle=\langle c_{j+}(t,x)c^\dagger_{k+}(0,0)\rangle^{1/2}\\
=&\sqrt{\frac{(i)^{j-k}\langle d_{+\rho}(t,x)d_{+,\rho}^\dagger(0,0)+ d_{+,\sigma}(t,x)d_{+,\sigma}^\dagger(0,0) \rangle}{2}} \\
=& \frac{\delta_{jk}}{\sqrt{2\pi i}}\prod_{\eta,\eta'=\pm} (u_{\eta\eta'}t-x-i0^+)^{-1/8}\ .
\end{split}
\end{equation}
Note that the semion correlation between different copies $j\neq k$ vanishes, which is consistent with the charge conservation in each copy of our model. In addition, we see that all the free fermion velocities $u_{\eta\eta'}$ contribute to the semion correlations, implying a fragmentation of the semions excitations.

\subsection{Exact solution for arbitrary $N$ and $M$}\label{sec:uniform-arbitraryNM}

We now extend our discussions to uniform interactions $J_{ij}=J$ with generic $N$ and $M$. In this case, the interaction term in Eq. (\ref{eq:Lint}) can be rewritten as
\beq\label{eq:Lint-Stot}
\mathcal{L}_{\text{int}}=-J\left(\bm{\mathcal{S}}_{\text{tot}} \cdot \bm{\mathcal{S}}_{\text{tot}}-\sum_{i=1}^N \bm{\mathcal{S}}_i \cdot \bm{\mathcal{S}}_i \right)\ ,
\eeq 
where we have defined $\bm{\mathcal{S}}_{\text{tot}}=\sum_{i=1}^N \bm{\mathcal{S}}_i$ as the total current. 

By the Sugawara construction\cite{DiFrancesco1997}, the term $\bm{\mathcal{S}}_i \cdot \bm{\mathcal{S}}_i$ in each copy is related to the boson field WZW Lagrangian by
\begin{equation}
\bm{\mathcal{S}}_i \cdot \bm{\mathcal{S}}_i=\frac{M+1}{8\pi^2}\sum_{\mu,\nu=1}^{M-1} K^{\mu\nu}_{SU(M)_1}\partial_x \phi_{i,\mu}\partial_x \phi_{i,\nu} \ .
\end{equation}
The term $J\sum_{i=1}^N\bm{\mathcal{S}}_i \cdot \bm{\mathcal{S}}_i$ together with $\mathcal{L}_0$ defined in Eq. (\ref{eq:L0}) are therefore equivalent to $N$ copies of decoupled chiral SU$(M)_1$ WZW models with a ``speed of light" $v_J=1-\frac{(M+1)J}{{2}\pi}$. 

On the other hand, the current operator $\bm{\mathcal{S}}_{\text{tot}}$ generates an $\mathfrak{su}(M)_N$ Kac-Moody algebra, as can be seen from Eq. (\ref{eq:KacMoody}) that the commutator $[\mathcal{S}^a_{\text{tot}}(x),\mathcal{S}^b_{\text{tot}}(x')]$ has an anomalous term $\frac{iN\delta^{ab}}{4\pi}\partial_x\delta(x-x')$. This leads us to take a decomposition of the full symmetry group $\otimes^N \text{SU}(M)_1$ (standing for the direct product of $N$ symmetry groups $\text{SU}(M)_1$) of the free part of our model into a subgroup $\text{SU}(M)_N$ (with current operator $\bm{\mathcal{S}}_{\text{tot}}$) and its coset: 
\beq\label{coset}
\begin{split}
\mathbb{G}_{M,N}=\left[\otimes^N \text{SU}(M)_1\right] /\text{SU}(M)_N\ .
\end{split}
\eeq
It is easy to see that the $\bm{\mathcal{S}}_{\text{tot}} \cdot \bm{\mathcal{S}}_{\text{tot}}$ term only contributes to the SU$(M)_N$ sector. The full Lagrangian therefore decomposes into two decoupled chiral models:
\begin{equation}\label{eq:Ltwoparts}
\mathcal{L}=\mathcal{L}^{\mathbb{G}_{M,N}}+\mathcal{L}^{\text{SU}(M)_N}\ ,
\end{equation}
where $\mathcal{L}^{\mathbb{G}_{M,N}}$ is the chiral $\mathcal{L}^{\mathbb{G}_{M,N}}$ coset theory with a central charge $c_{\mathbb{G}_{M,N}}=\frac{N(M-1)(N-1)}{N+M}$, and $\mathcal{L}^{\text{SU}(M)_N}$ is a chiral SU$(M)_N$ WZW model with a central charge $c_{\text{SU}(M)_N}=\frac{N(M^2-1)}{M+N}$. In particular, $\mathbb{G}_{M,2}$ is known to give the $\mathbb{Z}_M$ parafermion theory \cite{zam1985}. Since $-J\bm{\mathcal{S}}_{\text{tot}} \cdot \bm{\mathcal{S}}_{\text{tot}}$ term in Eq. (\ref{eq:Lint-Stot}) only contributes to $\mathcal{L}^{\text{SU}(M)_N}$, one can show the two parts in Eq. (\ref{eq:Ltwoparts}) have different speeds of light (see Appendix.~\ref{C2}):
\begin{equation}\label{eq:NMvelocities}
v_{\mathbb{G}_{M,N}}=1-\frac{(M+1)J}{2\pi},\ \  v_{\text{SU}(M)_N}=1+\frac{(N-1)J}{2\pi}.
\end{equation}
This agrees with the velocities we calculated for $N=2,M=2$ earlier in Eq. (\ref{eq:N=M=2-Majarana-velocity}) (when $J_\perp=J_z=J$). Again we require both velocities in Eq. (\ref{eq:NMvelocities}) to be positive, so that nonchiral reconstructions analogous to Fig. \ref{fig:largeK} do not happen. Since both models in Eq. (\ref{eq:Ltwoparts}) are exactly solvable in the conformal field theory (CFT), we conclude that our model with uniform interactions is integrable for any $N$ and $M$. Note that the chiral interaction of the model (Eq. (\ref{eq:Lint})) does break conformal symmetry, in the sense that it gives rise to two different ``speeds of light" in Eq. (\ref{eq:NMvelocities}).

As a side note, the strategy we adopted for solving $N=M=2$ case by performing suitable reorganizations of field operators is similar in spirit to those used in finding a Luther-Emery fixed point in the literature \cite{Kivelson1992,Luther1974,Emery1976}. To establish integrability, one can also contemplate mapping our model to equivalent known integrable models, such as integrable sine-Gordon or Gross-Neveu models\cite{Weiss1984,Babujian1999,Zamolodchikov1979,Witten1978,Shankar1985}. However, these known models are usually massive and involve both chirality.

\section{Random interaction: Quantum chaos in the large $N, M$ limit}\label{sec:randomcoupling}

We now turn to our model with random interactions $J_{ij}$ among the $N$ copies of $SU(M)_1$ chiral WZW models, and show that it is an exactly solvable quantum chaotic model in the large $N$ and $M$ limit. Note that here we assume $J_{ij}$ is random in the copy indices $i,j$, but is spatially uniform (namely, independent of position $x$).

In the same spirit as the 0+1d SY model \cite{SY93}, we assume the random interactions $J_{ij}=J_{ji}$ in the large $N$ limit ($M$ need not be large) have the following statistical means for copy indices $i<j$:
 \beq
 \begin{split}
 \la J_{ij} \ra =0\ ,\quad 
 \la J_{ij}J_{i'j'}\ra =\frac{J^2}{(N-1)(M-1)} \delta_{ii'} \delta_{jj'}\ ,
 \label{random}
 \end{split}
 \eeq 
where we define $J\ge0$ as the interaction strength. Such a scaling behavior regarding $N$ and $M$ ensures that the total energy density does not diverge in the large $N$ or $M$ limit for a fixed interaction strength $J$.

We first show that the model can be mapped into an interacting fermion model by enlarging its Hilbert space. To do so, we add to our model $\mathcal{L}$ in Eq. (\ref{SY}) the following free Lagrangian density of $N$ ancillary chiral boson (cb) fields $\phi'_{i,\rho}$ ($1\le i\le N$):
\begin{equation}\label{eq:Lancilla}
\mathcal{L}_{\text{ancilla}}^{\otimes^N \text{cb}}=-\frac{1}{4\pi}\sum_{i=1}^N\partial_x \phi'_{i,\rho}(\partial_t +\partial_x)\phi'_{i,\rho}\ ,
\end{equation}
which has the same ``speed of light" as that of the free part $\mathcal{L}_0$ in Eq. (\ref{eq:L0}) of our model. 
Since each copy $i$ of $\mathcal{L}_0$ in Eq. (\ref{eq:L0}) is a chiral SU$(M)_1$ WZW model, each copy $i$ of the sum $\mathcal{L}_0+\mathcal{L}_{\text{ancilla}}^{\otimes^N\text{cb}}$ can be embedded into a chiral U$(M)_1$ WZW model, which is known to have a free fermion representation. This allows us to map our model into a fermion model. First, for each copy $i$, there exists a linear recombination of $\phi_{i,\rho}'$ in Eq. (\ref{eq:Lancilla}) and $\phi_{i\mu}$ ($1\le\mu\le M-1$) in Eq. (\ref{eq:L0}) into $M$ new boson fields $\widetilde{\phi}_{i,\mu}$ ($1\le \mu\le M$) (see Appendix.~\ref{D}), such that
\begin{equation}\label{eq:L0+ancilla}
\mathcal{L}_0+\mathcal{L}_{\text{ancilla}}^{\otimes^N\text{cb}} =-\frac{1}{4\pi}\sum_{i=1}^N \sum_{\mu=1}^M\partial_x \widetilde{\phi}_{i,\mu}(\partial_t + \partial_x) \widetilde{\phi}_{i,\mu}\ ,
\end{equation}
and $\phi'_{i,\rho}=\frac{1}{\sqrt{M}}\sum_{\mu=1}^M \widetilde{\phi}_{i,\mu}$. 
We can then define a set of chiral fermion annihilation fields $f_{i,\mu}=e^{i\widetilde{\phi}_{i,\mu}}$, and one can prove that the SU$(M)_1$ current operators take the form
\begin{equation}
\mathcal{S}_i^a(t,x)=\sum_{\mu,\nu=1}^M f_{i,\mu}^\dag(t,x) T^a_{\mu\nu} f_{i,\nu}(t,x)\ ,
\end{equation}
where $T^a_{\mu\nu}$ ($1\le a\le M^2-1$) are generators of the fundamental representation matrix of the SU$(M)$ group . The full Lagrangian density with the ancillary fields can thus be rewritten into a chiral fermion model as
\begin{equation}\label{eq:Ltot-ancilla}
\mathcal{L}_{\text{tot}}=\mathcal{L}_0+\mathcal{L}_{\text{ancilla}}^{\otimes^N\text{cb}} +\mathcal{L}_{\text{int}}\ ,
\end{equation}
where the chiral fermion kinetic term is
\begin{equation}\label{eq:Ltot-ancilla1}
\mathcal{L}_0+\mathcal{L}_{\text{ancilla}}^{\otimes^N\text{cb}} =i\sum_{j=1}^N \sum_{\mu=1}^M f_{j,\mu}^\dag(\partial_t + \partial_x) f_{j,\mu}\ ,
\end{equation}
while the interaction term is
\begin{equation}
\mathcal{L}_{\text{int}} =-\sum_{i<j}^N J_{ij} \left[\sum_{\mu,\nu=1}^M f_{i,\mu}^\dag f_{i,\nu} f_{j,\nu}^\dag f_{j,\mu}-\frac{\hat{n}_i\hat{n}_j}{M}\right]\ ,
\end{equation}
with $\hat{n}_i=\sum_\mu f_{i,\mu}^\dag f_{i,\mu}$ defined as the total fermion density of copy $i$. Note that such a chiral fermion model in Eq. (\ref{eq:Ltot-ancilla}) with random $J_{ij}$ is analogous to the chiral SYK model of chiral Majorana fermions studied in Ref. \cite{Lian2019}, except that the model here consists of complex chiral fermions and has higher symmetries. This is also the natural $1+1$d generalization of the fermionic form of the $0+1$d SY model \cite{SY93}. 

We emphasize that due to the ancillary fields $\phi_{i,\rho}'$, the Hilbert space of the fermion model in Eq. (\ref{eq:Ltot-ancilla}) is enlarged compared to the original model in Eq. (\ref{SY}). Therefore, the fermions $f_{i,\mu}$ are subject to constraints analogous to those of the ``slave particles" in various quantum spin models\cite{Ioffe1989,Florens2004,Lee2005,Read_1983,Coleman1984,Kotliar1986}, and are not physical excitations. Instead, the physical charge excitations of our model in Eq. (\ref{SY}) are the Abelian anyons. However, as we will show below, all the anyon correlations can be computed from the fermion correlations and the ancillary boson correlations.

\subsection{The anyon two-point function}

The annihilation operator for the SU$(M)_1$ Abelian anyon with minimal charge  in each copy of our model in Eq. (\ref{SY}) can be written in terms of the ancillary boson field $\phi_{i,\rho}'$ and the fields $\widetilde{\phi}_{i,\mu}$ we introduced in Eq. (\ref{eq:L0+ancilla}) as 
\begin{equation}\label{eq:anyon-def}
\chi_i=e^{i\widetilde{\phi}_{i,1}-i\frac{\phi'_{i,\rho}}{\sqrt{M}}}\ ,
\end{equation}
which has a scaling dimension $\Delta=\frac{M-1}{2M}$ and statistical phase $2\pi\Delta$. All the other Abelian anyons in the system can be obtained from fusions of this anyon in Eq. (\ref{eq:anyon-def}).

In the enlarged model of Eq. (\ref{eq:Ltot-ancilla}), the ancillary boson field $\phi'_{i,\rho}$ only contributes to $\mathcal{L}_{\text{ancilla}}^{\otimes^N\text{cb}}$, and is decoupled with the remaining part $\mathcal{L}_0+\mathcal{L}_{\text{int}}$. Therefore, its correlation is entirely determined by $\mathcal{L}_{\text{ancilla}}^{\otimes^N\text{cb}}$ in Eq. (\ref{eq:Lancilla}) as
\begin{equation}\label{eq:phirho-correlation}
\langle \phi'_{i,\rho}(t,x)\phi'_{j,\rho}(0,0)\rangle =-\delta_{ij}\log [2\pi i(t-x-i0^+)]\ ,
\end{equation}
up to constant shifts of $\phi'_{i,\rho}$. On the other hand, since $\phi'_{i,\rho}=\frac{1}{\sqrt{M}}\sum_{\mu=1}^M \widetilde{\phi}_{i,\mu}$, by expressing $\widetilde{\phi}_{i,\mu}$ in terms of $\phi'_{i,\rho}$ (see Appendix.~\ref{D} for details), one can show that 
\begin{equation}\label{eq:phirho-correlation2}
\langle \widetilde{\phi}_{i,\mu}(t,x)\phi'_{j,\rho}(0,0)\rangle=\frac{1}{\sqrt{M}}\langle \phi'_{i,\rho}(t,x)\phi'_{j,\rho}(0,0)\rangle
\end{equation}
for any $1\le\mu\le M$. These relations allow us to express the correlations of anyons in Eq. (\ref{eq:anyon-def}) as functions of the fermion correlations of the fermion model in Eq. (\ref{eq:Ltot-ancilla}).

In this section, we calculate the following averaged anyon two-point function as an example:
\begin{equation}
G_\chi(t,x)=\frac{1}{N}\sum_{i=1}^N\langle \chi_i(t,x)\chi^\dagger_i (0,0) \rangle\ .
\end{equation}
By Eq. (\ref{eq:anyon-def}), we can rewrite the anyon two-point function as
\begin{equation}\label{eq:anyon-correlation-fermion}
\begin{split}
&G_{\chi}(t,x)=\frac{1}{N}\sum_{i=1}^N\langle e^{i\left(\widetilde{\phi}_{i,1}-\frac{\phi'_{i,\rho}}{\sqrt{M}}\right)(t,x)}e^{-i\left(\widetilde{\phi}_{i,1}-\frac{\phi'_{i,\rho}}{\sqrt{M}}\right)(0,0)} \rangle\\
&=\frac{1}{N}\sum_{i=1}^N e^{\frac{1}{M}\langle \phi'_{i,\rho}(t,x) \phi'_{i,\rho}(0,0) \rangle}e^{-\frac{1}{\sqrt{M}}\langle \widetilde{\phi}_{i,1}(t,x)\phi'_{i\rho}(0,0)\rangle}\\ &\qquad \times e^{-\frac{1}{\sqrt{M}}\langle \phi'_{i,\rho}(t,x)\widetilde{\phi}_{i,1}(0,0)\rangle}  \langle e^{i\widetilde{\phi}_{i,1}(t,x)}e^{-i\widetilde{\phi}_{i,1}(0,0)} \rangle\\
&=[2\pi i(t-x-i0^+)]^{1/M} G_f(t,x)\ ,
\end{split}
\end{equation}
where we have defined the fermion two-point function of the model in Eq. (\ref{eq:Ltot-ancilla}):
\begin{equation}\label{eq:fermion-correlation}
G_f(t,x)=\frac{1}{N}\sum_{i=1}^N\langle f_{i,1}(t,x)f^\dagger_{i,1} (0,0) \rangle\ ,
\end{equation}
and we have used the boson correlations in Eqs. (\ref{eq:phirho-correlation}), (\ref{eq:phirho-correlation2}) and the definition of the fermion fields $f_{i,\mu}=e^{i\widetilde{\phi}_{i,\mu}}$ in Eq. (\ref{eq:Ltot-ancilla}). Therefore, we see from Eq. (\ref{eq:anyon-correlation-fermion}) that the anyon correlation decomposes into a product of two parts: a vertex correlation of the U(1) ancillary boson fields which governs the anyon fractional statistics, and a fermion correlation of the interacting fermion model in Eq. (\ref{eq:phirho-correlation}). Note that this decomposition holds for any $N$ and $M$. Expressions similar to Eq. (\ref{eq:anyon-correlation-fermion}) can also be derived for any other anyon correlations (e.g., see Sec. \ref{sec:randomOTOC} below).

\begin{figure}[tbp]
\centering\includegraphics[width=0.5\textwidth]{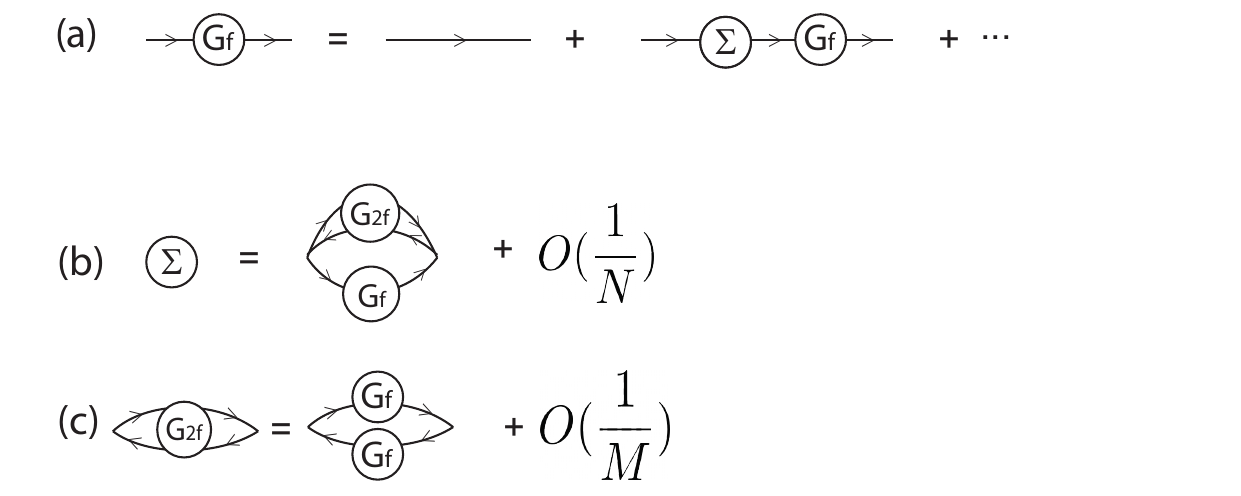}
\caption{Melon diagrams contributing to the two point correlation function $G_f$ in the large $N,M$ limit. (a) $G_f$ is the single fermion propagator, and $\Sigma$ is the self-energy function. (b) Feynman diagram representation of the relations between self-energy $\Sigma$ and two-point functions $G_f,G_{2f}$ following the Schwinger-Dyson equation. (c) The relation between $G_f$ and $G_{2f}$, the double fermion propagator of fermion bilinear operators $f_{j,\mu}^\dag f_{j,\nu}$.
}\label{fig:melon}
\end{figure}

With random interactions $J_{ij}$ satisfying Eq. (\ref{random}), the fermion two-point function in Eq. (\ref{eq:fermion-correlation}) is exactly solvable in the large $N$ and $M$ limit by techniques employed in the SY model \cite{SY93} and the SYK models \cite{Polchinski:2016xgd,Maldacena:2016hyu,Kitaev:2017awl,Lian2019}. In the large $N$ and large $M$ limit, one can prove that the leading order contributions to the two-point fermion correlation are the melon diagrams as shown in Fig. \ref{fig:melon} (see also Fig. \ref{fig:diag} in Appendix.~\ref{E}). This leads to the same Schwinger-Dyson (SD) equation appeared in the $1+1$d chiral SYK model studied in Ref. \cite{Lian2019}, as given in Eq. (\ref{seq:SD}). The fermion two-point correlation in the large $N$ and $M$ limit is then equal to:

\beq\label{eq:Gf}
G_f(t,x)=\frac{1}{2\pi i\sqrt{(u_+ t-x-i0^+)(u_- t-x-i0^+)}}\ ,
\eeq 
where the velocities are defined by
\begin{equation}
u_{\pm}=1\pm\frac{J}{2\pi}\ .
\end{equation}
When $J>2\pi$, the chirality of the $u_-$ mode will not be preserved ($u_-<0$), and mode reconstructions similar to that in Fig. \ref{fig:largeK}(b) need be considered. In this paper, we require 
\begin{equation}
0\le J< 2\pi \ ,
\end{equation}
so that $u_\pm>0$ and the chirality of the model is preserved. This yields the anyon correlation function in the large $N$ and $M$ limit:
\beq\label{eq:anyon-correlation-final}
G_\chi(t,x)=\frac{(2\pi i)^{-1+\frac{1}{M}}(t-x-i0^+)^{\frac{1}{M}}}{\sqrt{(u_+ t-x-i0^+)(u_- t-x-i0^+)}}\ .
\eeq 
Note that when $M\rightarrow\infty$, the anyon correlation tends to the fermion correlation, which is because the anyon fractional statistics tends to the fermion statistics in the large $M$ limit. For finite $M$ (while $N\rightarrow\infty$), we expect the anyon two-point correlation in Eq. (\ref{eq:anyon-correlation-final}) to acquire an order $O(1/M)$ correction (see Fig.~\ref{fig:diag} (b)).

At finite temperature $\beta^{-1}$, one can show that anyon two-point function $G_\chi^\beta(t,x)$ given in Eq. (\ref{eq:anyon-correlation-final}) are modified using the ``conformal transformation"
\begin{equation}
u_{\eta}t-x-i0^+ \rightarrow \frac{u_{\eta}\beta}{\pi} \sinh \frac{\pi}{\beta}(t-u_{\eta}^{-1}x-i0^+)
\end{equation} for all velocities $u_\eta\in \{u_\pm$,$1$\}, although the $1+1$d interacting chiral model we studied does not have the full conformal symmetry (as evident from the presence of multiple ``speed of light").

\subsection{The anyon out-of-time-ordered correlation: quantum chaos}\label{sec:randomOTOC}

We now show that our model in Eq. (\ref{SY}) with random interactions $J_{ij}$ is quantum chaotic in the large $N$ and $M$ limit. One of the evidences of quantum chaos is the presence of a positive Lyapunov exponent in the finite temperature out-of-time-order correlation (OTOC), which resembles the Lyapunov exponent of classical chaos. We define the regularized OTOC for the anyons in our model at temperature $\beta^{-1}$ as
\begin{equation}
\begin{split}
&\mathcal{F}_\chi(t_1,x_1;t_2,x_2)\equiv\\ &\frac{1}{N^2}\sum_{i,j}^N \text{Tr}\left[y\chi_{j}^\dagger(t_1,x_1)y\chi_{i}(0,0)y\chi_{j}(t_2,x_2)y\chi_{i}^\dagger(0,0)\right],
\end{split}
\end{equation}
where $y=e^{-\beta H/4}$ split the four anyon fields evenly by one quarter of the thermal circle ($H$ is the Hamiltonian). Such a splitting of the partition function $e^{-\beta H}$  makes the OTOC $\mathcal{F}_\chi(t_1,x_1;t_2,x_2)$ a real function, and avoids its divergence when $(t_j,x_j)\rightarrow (0,0)$. This regularized OTOC shares the same growing behavior as the unregularized OTOC.

Following the anyon expression in Eq. (\ref{eq:anyon-def}) and a derivation similar to Eq. (\ref{eq:anyon-correlation-fermion}), we can express the anyon OTOC in terms of the fermion correlations up to order $O(\frac{1}{NM})$ as
\begin{equation}\label{eq:otoc-expansion}
\begin{split}
&\mathcal{F}_\chi(t_1,x_1;t_2,x_2)=\mathcal{F}^0_\chi(t_1,x_1;t_2,x_2) \\
&\qquad\quad -\frac{1}{NM}\mathcal{C}(t_1,x_1;t_2,x_2)\delta \mathcal{F}(t_1,x_1;t_2,x_2)\ ,
\end{split}
\end{equation}
where 
\begin{equation}
\mathcal{F}^0_\chi(t_1,x_1;t_2,x_2)=G^\beta_\chi(-i\frac{\beta}{2},0) G^\beta_\chi(t_{21}-i\frac{\beta}{2},x_{21})
\end{equation}
is a non-growing piece from the anyon two-point functions $G^\beta_\chi$ within each copy, with $t_{21}=t_2-t_1$ and $x_{21}=x_2-x_1$ defined;
\begin{equation}\label{eq:Cfactor}
\mathcal{C}(t_1,x_1;t_2,x_2)=\left[4i\beta^2\sinh{\frac{\pi}{\beta}(t_{21}-i\frac{\beta}{2}-x_{21})} \right]^{1/M}
\end{equation}
is the contribution from vertex correlations of the ancillary boson fields $\phi_{i,\rho}'$, while $\delta \mathcal{F}(t_1,x_1;t_2,x_2)$ is a fermion four-point function given by an infinite fermion ladder Feynman diagram summation (see Appendix.~\ref{F2}). In general, the factor $\mathcal{C}(t_1,x_1;t_2,x_2)$ in Eq. (\ref{eq:Cfactor}) has only order $O(\frac{1}{N})$ correction terms, which is thus exact for any $M$ in the large $N$ limit. In contrast, the fermion correlation has both order $O(\frac{1}{N})$ and $O(\frac{1}{M})$ corrections.

\begin{figure}[tbp!]
\centering\includegraphics[width=0.5\textwidth]{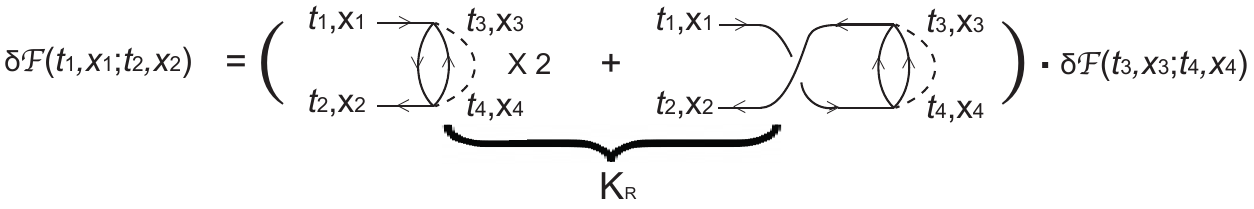}
\caption{Feynman diagram illustration that the OTOC function $\delta F$ is an eigenfunction of the kernel $K_R$.}\label{fig:ladder}
\end{figure}

The infinite sum of $\delta \mathcal{F}(t_1,x_1;t_2,x_2)$ can be incorporated into a self-consistent equation:

\beq
\begin{split}
\delta\mathcal{F}(t_1,x_1;t_2,x_2)=&\int dt_3dt_4dx_3dx_4 K_R(t_1,\cdots,x_4)\\
&\times \delta\mathcal{F}(t_3,x_3;t_4,x_4)\ ,
\end{split}
\eeq 
where the kernel $K_R$ is illustrated in Fig. \ref{fig:ladder} and explicitly given in Eq. (\ref{seq:KR}). 

As shown in Appendix.~\ref{F2}, for this self-consistent equation, the key conclusion is, in the Lyapunov regime of time $\beta <t<\beta\log (NM)$, the OTOC function $\delta\mathcal{F}$ grows in time with a positive velocity dependent Lyapunov exponent (VDLE) $\lambda_v$ within the causality cone of the model, which therefore indicates quantum chaos. 
The VDLE is defined as follows. Taking $x_1=x_2=x$ and $t_1=t_2=t$, we define the reduced OTOC function $\delta\mathcal{F}(t,x)=\delta\mathcal{F}(t,x;t,x)$. Note that the factor $\mathcal{C}(t_1,x_1;t_2,x_2)$ is then a constant independent of $t$ and $x$, since it only depends on $t_{21}$ and $x_{21}$. Along a fixed velocity $v=x/t$, this OTOC function is expected to grow as
\beq\label{grow}
\delta\mathcal{F}(t,vt) \sim  e^{\lambda_v t}
\eeq 
within the Lyapunov regime $\beta <t<\beta\log (NM)$, where $\lambda_v$ is defined as the VDLE. When $t>\beta\log (NM)$, the higher order terms in the $\frac{1}{N}$ and $\frac{1}{M}$ expansions in Eq. (\ref{eq:otoc-expansion}) become non-negligible, and the growth of the order $\frac{1}{NM}$ piece $\delta\mathcal{F}$ is no longer meaningful. We find the VDLE $\lambda_v$ of the OTOC is positive for velocities within $u_-<v<u_+$ (recall that $u_\pm=1\pm\frac{J}{2\pi}$), which is given by
\begin{equation}\label{eq:Lv}
\lambda_v=
\begin{cases}
& \frac{2\pi}{\beta}(vu_-^{-1}-1)\ ,\ \qquad (u_-<v<v_*') \\
& \overline{\lambda}-\frac{2\pi\xi}{\beta}(v-v_c)^2 \ ,\quad (v_*'<v<v_*) \\
& \frac{2\pi}{\beta}(1-vu_+^{-1})\ ,\ \qquad (v_*<v<u_+) \\
\end{cases}
\end{equation}
where
\beq\label{eq:maxVDLE}
\overline{\lambda}=\lambda_v^{\text{max}}=\frac{2\pi}{\beta} \varkappa(0)=\frac{2\pi}{\beta}\frac{\mathcal{J}(\sqrt{3-2\mathcal{J}^2}-1)}{1-\mathcal{J}^2}
\eeq 
is the maximal VDLE, with $\mathcal{J}\equiv\frac{J}{2\pi}$. This maximal VDLE is reached at velocity $v_c=1-\frac{\mathcal{J}^2}{\sqrt{3-2\mathcal{J}^2}}$. The other parameters are given by $v_*=\frac{2-2\mathcal{J}^2}{2-\mathcal{J}}$, $v_*'=\frac{2-2\mathcal{J}^2}{2+\mathcal{J}}$, and $\xi=\frac{(3-2\mathcal{J}^2)^{3/2}}{6\mathcal{J}(1-\mathcal{J}^2)^2}$. Note that $u_-$ and $u_+$ give the causality velocity boundary of the fermion retarded Green's function $G_R(t,x)=2\Theta(t)\text{Re}[G_f(t,x)]$, with $G_f$ defined in Eq. (\ref{eq:Gf}) and $\Theta(t)$ the Heaviside step function. Therefore, a positive VDLE $\lambda_v$ within the velocity range $[u_-,u_+]$ indicates quantum chaos of all the propagating information of the chiral system. The results in Eqs. (\ref{eq:Lv}) and (\ref{eq:maxVDLE}) share similarity with that of the chiral SYK model in Ref. \cite{Lian2019}, except that the OTOC here are for anyons.

\begin{figure}[tbp]
\centering\includegraphics[width=0.45\textwidth]{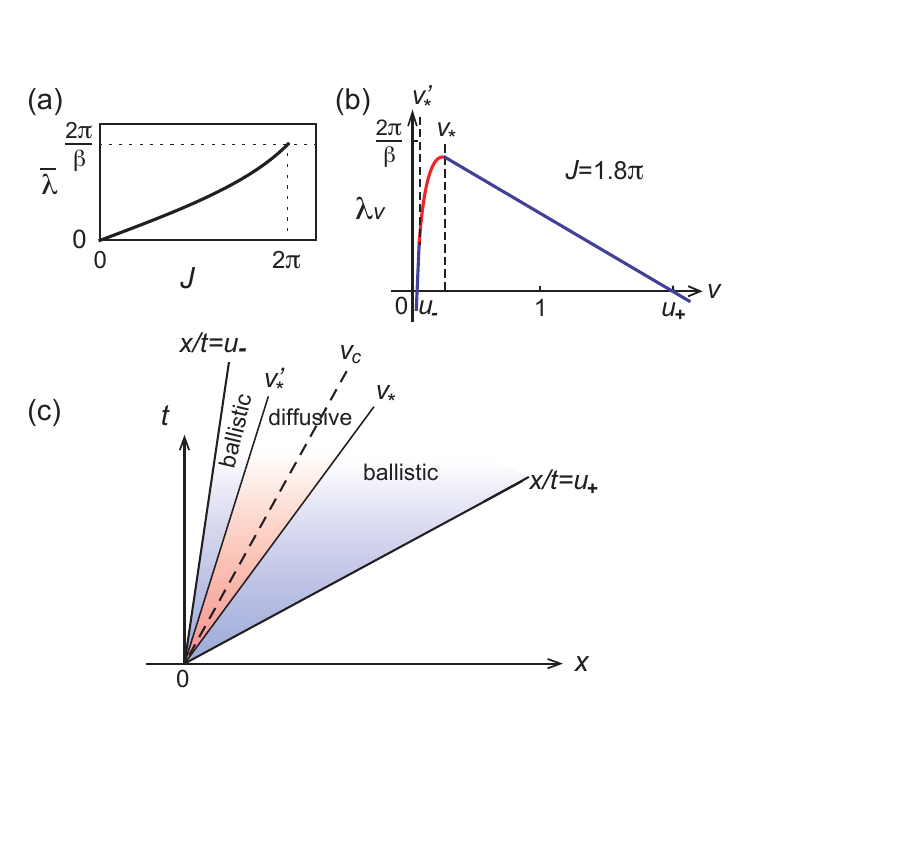}
\caption{(a) Plot of the maximal velocity-dependent Lyapunov exponent (VDLE) $\lambda_\nu^{\text{max}}=\overline{\lambda}$ as a function of $0\leq J<2\pi$. The Lyapunov exponent saturates the maximum value $\frac{2\pi}{\beta}$ as $J\rightarrow 2\pi$. (b) The velocity-dependent Lyapunov exponent (VDLE) $\lambda_\nu(v)$ as a function of velocity for $J = 1.8\pi$. $u_-<v<v_*'$ and $v_*<v<u_+$ are the ballistic regime (blue curve) where the relation is linear in $v$. $v_*'<v<v_*$ is the diffusive regime (red curve) where the relation is quadratic in $v$. The VDLE reaches its maximum at velocity $v_c$. (c) The OTOC $\delta\mathcal{F}$ grows exponentially in time in the regime between velocities $u_-$ and $u_+$ with the VDLE $\lambda_\nu(v)$ in (b).}
\label{fig:butt}
\end{figure}

Fig. \ref{fig:butt} (a) shows the maximal VDLE $\overline{\lambda}$ in Eq. (\ref{eq:maxVDLE}) as a function of $J$. In particular, $\overline{\lambda}$ reaches the maximal quantum chaos exponent bound $2\pi/\beta$ when $J\rightarrow 2\pi$, i.e., the maximal interaction strength preserving the chirality of the model. As shown in Eq. (\ref{eq:Lv}) and Fig.~\ref{fig:butt} (b) and (c), the VDLE has a linear dependence on the velocity $v$ in regimes with $u_-<v<v_*'$, $v_*<v<u_+$. These are known as the ballistic regimes of VDLE where $\lambda_v$ saturates its velocity-dependent upper bound \cite{Mezei2020,Chowdhury2017}. In contrast, the VDLE has a quadratic velocity $v$ dependence in the regime $v_*'<v<v_*$, which is known as the diffusive regime, and the maximal VDLE is reached in this regime (at velocity $v_c$) \cite{Mezei2020,Chowdhury2017}.

From Eq. (\ref{eq:otoc-expansion}), we again see that the fractional statistics of Abelian anyons in the growing piece of the OTOC is encoded in the factor $\mathcal{C}(t_1,x_1;t_2,x_2)$, which is solely contributed by the U(1) ancillary boson fields $\phi_{i,\rho}'$. The remaining factor $\delta\mathcal{F}$ is the usual OTOC growing piece for complex fermions measuring the onset many-body quantum chaos characterized by the Lyapunov exponent. Although our calculations here are only accurate to order $\frac{1}{NM}$, we conjecture that the quantum chaos persists to finite $N$ and $M$ beyond a threshold, due to the nonlinear nature of our model with random interactions. This threshold should be above $N=M=2$, which has only one interaction and is integrable as we showed earlier.

\section{Discussion}\label{sec:discussion}

We have studied the quantum integrability and chaos of a $1+1$d chiral SY model consisting of N copies of the SU$(M)_1$ chiral WZW models with chiral current-current interactions among each other with coefficients $J_{ij}$ ($i\neq j$), which host Abelian anyons as charge excitations. This model gives a minimal $1+1$d generalization of the $0+1$d SY model \cite{SY93}, a spin model of $N$ coupled SU($M$) spins. We have shown that the $1+1$d chiral SY model we studied is integrable for any $N$ and $M$ when the interactions $J_{ij}$ are uniform (independent of indices $i$ and $j$), In contrast, it is quantum chaotic in the large $N$ and $M$ limit when the interactions $J_{ij}$ are random in indices $i$ and $j$. Particularly, we are able to investigate the quantum chaos of $1+1$d anyons with such a model. In physical systems such as the edges of FQH systems, integrability or chaos are expected to significantly affect the interference of the edge states in edge interferometers such as the Fabry-P\'erot geometry \cite{Bartolomei2020,Carrega2021,McClure2012,Ofek2010,Halperin2011,Fu2009,Akhmerov2009,Lian2018}.

In both the $N=M=2$ integrable case and the chaotic case with random interactions in the large $N$, $M$ limit, we derived the two-point correlation functions of the Abelian anyons with the lowest scaling dimension. Intriguingly, the two-point functions of these two cases behave similarly, exhibiting a fractionalization into different characteristic velocities. However, their four-point functions are drastically different. We showed that the anyon OTOC four-point function of the large $N$ and $M$ limit with random interactions has a positive Lyapunov exponent similar to that of the chiral SYK model studied in Ref. \cite{Lian2019}, as expected for quantum chaos. The maximal VDLE approaches the maximal chaos bound $2\pi/\beta$ \cite{Shenker2014,Shenker2015,Maldacena:2015waa} when the interaction strength $J$ reaches its physical upper bound $2\pi$. In contrast, the integrable case with uniform interactions cannot have a positive Lyapunov exponent \cite{Lian2019}, due to the solvable eigen-spectrum of the model.

Unlike the 0+1d SY model and SYK model which have a large zero-temperature entropy reflecting their ``spin glass" like nature \cite{SY93,Polchinski:2016xgd,Maldacena:2016hyu,Kitaev:2017awl}, the 1+1d chiral SY model here and the chiral SYK model in Ref. \cite{Lian2019} have a vanishing zero-temperature entropy density for $0\le J<2\pi$ in the quantum chaos regime. This is because as long as the chirality of the model is preserved, any excitations will be energetically lower bounded by the velocity $u_-=1-\frac{J}{2\pi}$ times their momenta, which avoids an exponentially large density of states at zero energy.

The chiral interactions in our chiral SY model preserves the scaling symmetry but break conformal symmetry. This leads to the absence of a unique ``speed of light" in the model. As a result, in the integrable regime, the model decomposes into CFT theories with different ``speeds of light"; while in the chaotic regime, the model exhibit a chiral causality cone of a range of velocity ($[u_-,u_+]$ as shown in Fig. \ref{fig:butt}(c)). We note that the chiral causality cone shown in Fig. \ref{fig:butt}(c) resembles the causality cone of a 1+1d nonchiral CFT model in a reference frame moving faster than the speed of light, so that all the information propagate in the same direction. Therefore, we speculate the intrinsic analogy between the chiral SY model here (as well as the chiral SYK model in Ref. \cite{Lian2019}) and the quantum chaotic nonchiral CFTs with large central charges \cite{Turiaci_2016}. In addition, although the chirality and the absence of conformal symmetry makes it difficult to find a bulk gravity dual of quantum chaotic chiral models studied here, it is possible that their gravity dual, if exists, lives in a 2+1d spacetime that rotates faster than the speed of light at its boundary.

The correlation functions of the 0+1d complex fermion SYK model are known to have a chemical potential dependence (e.g., see \cite{Gu2020} and references therein). In our case, the 1+1d chiral SY model in its enlarged complex fermionic form of Eq. (\ref{eq:Ltot-ancilla}) has a much simpler chemical potential dependence: adding a chemical potential term $\epsilon_F\sum_{j,\mu}f^\dag_{j,\mu}f_{j,\mu}$ (which preserves the SU($M$) global symmetry) is equivalent to doing a spatially dependent unitary transformation $f_{j,\mu}(x)\rightarrow e^{i\epsilon_F x} f_{j,\mu}(x)$, or equivalently, a boson field transformation $\widetilde{\phi}_{j,\mu}(x)\rightarrow \widetilde{\phi}_{j,\mu}(x)+\epsilon_F x$. Accordingly, all correlation functions can be derived simply by the above unitary transformation.

In calculating the anyon correlation functions, a useful technique we have employed is the mapping of the systems with anyon excitations into fermion models with enlarged Hilbert space, which we summarize here for applications in more generic $1+1$d chiral models (FQH edge theories). There are two different situations where such mappings exist:

(1) First, if a $1+1$d model can be mapped into free Majorana fermions (thus is integrable), in analogy to the $N=M=2$ integrable example in Sec. \ref{sec:N=2M=2freeMajorana-0}, we expect that the anyon correlations can be exactly solved by adding a duplicate copy of the model. This is because two identical Majorana fermion modes are equivalent to a complex fermion mode, which can be bosonized to establish a connection with the anyon vertex operators. Moreover, in such calculations, the SO(8) triality mapping provides a key transformation from the $p$-wave pairing terms (which yield Majorana fermions) into four fermion interaction terms, which can be further transformed into free boson terms.

(2) The second situation is regardless of integrability: Generically, the action of a $1+1$d model with Abelian anyons can be written in terms of a set of boson fields $\bm{\phi}=(\phi_1,\cdots,\phi_r)^T$. Assume by adding a set of free ancillary boson fields $\bm{\phi}'=(\phi'_1,\cdots,\phi'_s)^T$, all the boson fields $\bm{\phi}$ and $\bm{\phi}'$ can be linearly recombined into a new basis of boson fields $\widetilde{\bm{\phi}}=(\widetilde{\phi}_1,\cdots,\widetilde{\phi}_{r+s})^T=Q(\bm{\phi}'^T,\bm{\phi}^T)^T$ with compactification radius $1$ (namely,the $K$ matrix is identity), such that the enlarged model with ancillary fields maps to a fermion model of fermion modes $f_i=e^{i\widetilde{\phi}_i}$. Here the transformation matrix $Q$ is a real nonsingular matrix. Then, for all anyon operators of the form 
\begin{equation}\label{eq:generic-anyons}
\chi_\alpha=e^{i\bm{l}_\alpha\cdot\widetilde{\bm{\phi}}+i\bm{p}_\alpha\cdot\bm{\phi}'}\ ,
\end{equation}
where $\bm{l}_\alpha$ is a length $r+s$ vector with integer entries and $\bm{p}_\alpha$ is a length $s$ vector of real numbers, their anyon correlation functions can be calculated from the fermion correlation functions. More concretely, the correlation function of $n$ anyon operators factorizes into the following two parts:
\beq\label{gc}
\begin{split}
&\langle \prod_{j=1}^n\chi_{\alpha_j}(t_j,x_j) \rangle=\mathcal{C}(t_i,x_j)\mathcal{D}_f(t_i,x_j)\ .
\end{split}
\eeq
The first factor is defined as follows and only involves the correlation functions of the free ancillary fields $\bm{\phi}'$, which is easily calculable:
\begin{equation}\label{gc1}
\begin{split}
&\mathcal{C}(t_i,x_j)=e^{-\sum_{i\neq j}^n  \langle \left[\frac{\bm{p}_{\alpha_i}}{2}\cdot \bm{\phi}'(t_i,x_i) + \boldsymbol{l}_{\alpha_i} \cdot \widetilde{\boldsymbol{\phi}}(t_i,x_i)\right] \bm{p}_{\alpha_j}\cdot \bm{\phi}'(t_j,x_j)  \rangle}\\
&=\prod_{1=i<j}^n [2\pi i (t_{ij}-x_{ij}-i0^+)]^{-\bm{p}_{\alpha_i}\cdot \bm{p}_{\alpha_j} -\boldsymbol{l}_{\alpha_i}^T \widetilde{Q}\bm{p}_{\alpha_j} -\boldsymbol{l}_{\alpha_j}^T \widetilde{Q}\bm{p}_{\alpha_i} },
\end{split}
\end{equation}
where $(t_{ij},x_{ij})=(t_i-t_j,x_i-x_j)$, and $\widetilde{Q}$ is the first $s$ columns of matrix $Q$. With $\bm{p}_{\alpha_i}$ generically not integer valued, this factor reflects the fractional statistics of the Abelian anyons. The second factor is
\begin{equation}\label{gc2}
\mathcal{D}_f(t_j,x_j)=\langle \prod_{j=1}^n e^{i\boldsymbol{l}_{\alpha_j} \cdot \widetilde{\boldsymbol{\phi}}(t_j,x_j)}\rangle\ ,
\end{equation}
which is equivalent to an $n$-point fermion correlation. This can be seen by mapping the vertex operators into fermions as $e^{i\widetilde{\phi}_j}\rightarrow f_j$, $e^{2i\widetilde{\phi}_j}\rightarrow -\frac{i}{2\pi}f_j\partial_x f_j$, $e^{3i\widetilde{\phi}_j}\rightarrow \frac{i}{16\pi^3}f_j\partial_x f_j \partial_x^2 f_j$, and so on. This reduces the calculation of $\mathcal{D}_f(t_j,x_j)$ into a fermion problem (though usually interacting), which is generically easier, since the Feynman rules for fermions are much simpler than those for the anyon vertex operators in their original representation. Moreover, the numerical calculations (exact diagonalization, etc) of fermion models are also much more straightforward than those of models in the anyon representations.

We emphasize that not all the $1+1$d models with Abelian anyons can be mapped into fermion models by the above method of adding ancillary boson fields, but the chiral SY model we studied here belongs to this class. With the free ancillary boson fields of Eq. (\ref{eq:Lancilla}) added to the chiral SY model (for any interactions $J_{ij}$ and any $N$ and $M$), all the Abelian anyons take the form of Eq. (\ref{eq:generic-anyons}). In this paper, we have restricted our attention to the correlations of anyons with the vector $\bm{l}_\alpha$ satisfying $|\bm{l}_\alpha|=1$ as given in Eq. (\ref{eq:anyon-def}), which are the ``elementary anyons" in the fusion rules. Correlations of anyons with $|\bm{l}_\alpha|>1$ will involve the correlations of bilinears or higher order products of fermion operators, as is clear from Eqs. (\ref{gc})-(\ref{gc2}). The calculation of these higher order correlations is, however, more complicated and beyond the scope of this paper. 

An intriguing future question is to generalize our study of quantum chaos into the 1+1d interacting Abelian anyon models which cannot be embedded into fermion models, and further into $1+1$d models hosting non-Abelian anyons, in which case our method presented in this paper may not apply. This calls for the exploration of solvable limits regarding the Feynman rules of vertex operators and other anyonic operators, which has not been studied yet and is challenging. A simple generalization of our model here would be to consider $N$ copies of other chiral WZW models with current-current interactions, such as the chiral O$(M)_1$ WZW model, which has a Majorana fermion representation for its currents and contains non-Abelian anyons if $M$ is odd. In particular, the non-local nature of non-Abelian anyons such as the Ising anyons \cite{moore1991} and the Fibonacci anyons \cite{hu_2018}, as indicated by their irrational Hilbert space dimensions and fusion structures, may yield additional stringent constraints on their correlations in the presence of quantum chaos. It would also be interesting to examine the competition among different chiral interaction terms of the same scaling dimension, which have their scaling dimensions fixed by their conformal spins and thus can hardly dominate over one another under renormalization group flows.

\section*{Acknowledgements}
The authors thank Zhenbin Yang, Fabian Essler, Jeffrey C.Y. Teo and Igor Klebanov for helpful discussions. Y.H. is supported by grant EP/S020527/1 from EPSRC. B.L. acknowledges support from the Alfred P. Sloan Foundation. This work is also supported by NSF through the Princeton University’s Materials Research Science and Engineering Center DMR-2011750.

\newpage

\appendix
\onecolumngrid
\section{K-matrix for $M-1$~-component $(2,2,1)$ Halperin states}\label{A}
In this appendix, we establish the equivalence between the edge theory for a $M-1$-component $(2,2,1)$ Halperin state and $SU(M)_1$ WZW conformal field theory in Eq. ~(\ref{eq:L0}).

The Lagrangian density of a $M-1$-component $(2,2,1)$ Halperin state\cite{Blok1992} is
\begin{equation}
\mathcal{L}_{(2,2,1)}=-\frac{1}{4\pi} \sum_{i=1}^N\sum_{\mu,\nu=1}^{M-1}K_{M (2,2,1)}^{\mu\nu}\partial_x \bar{\phi}_{i\nu}( \partial_t +\partial_x) \bar{\phi}_{i\mu}\ ,
\end{equation}
where 
\beq
K_{M (2,2,1)}=\begin{pmatrix}
2      & 1      & 1 & \dots  & 1 & 1 \\
1      & 2      & 1 & \dots  & 1 & 1 \\
1      & 1      & 2 &        &   & 1 \\
\vdots & \vdots &   & \ddots &   & \vdots \\
1      & 1      &   &        & 2 & 1 \\
1      & 1      & 1 & \dots  & 1 & 2
\end{pmatrix}
\eeq with charge vector $\bar{q}=(1,1,\cdots,1)^T$

at filling $\nu=\frac{M-1}{M}$. We can transform the above K-matrix into the Cartan matrix for SU$(M)$ (i.e., the K-matrix in Eq. (\ref{eq:L0})) by the following basis transformation $K_{SU(M)}=G^T K_{M (2,2,1)} G$:
\beq
G=\begin{pmatrix}
1     & -1       \\
 & 1      & -1\\
     &     & \ddots & \ddots      \\
     &  &   &  &1   & -1 \\
     &      &   &        & & 1 
\end{pmatrix}\ ,
\eeq with charge vector $q=G\bar{q}=(0,\cdots,0,1)^T$.

\section{The point-splitting procedure}\label{app:B}
In this appendix, we derive the point-splitting procedure along the $x$-direction (as is conventional in condensed matter systems) for fermion fields via bosonization.
Let us start by rewriting a bosonic field $\phi(x)$ of radius $1$ (on a constant time slice) as
\be
\phi(x)=\varphi(x)+\varphi^\dagger(x)\ ,
\ee where $\varphi(x), \varphi^\dagger(x)$ represents the annihilation and creation operators in the mode expansion of $\phi(x)$, and satisfy $[\varphi(x),\varphi(x')]=[\varphi^\dagger(x), \varphi^\dagger(x')]=0$.  Consistent with the commutation relation
\beq
[\partial_x\phi(x),\phi(x')]=2\pi i \delta(x-x')\ ,
\eeq we have
\beq
[\varphi(x),\varphi^\dag(x')]=-\log[-2\pi i(x-x'+i0^+)]\ .
\eeq

Define the fermion creation and annihilation operators $c(x)=e^{i\phi(x)}$ and $c^\dagger(x)=e^{-i\phi(x)}$. With a point splitting in the $x$-direction, we can now use normal ordering to calculate their operator product expansions (OPE). The normal ordering will always put $\varphi^\dagger(x)$ to the left side of $\varphi(x)$, and we use $\normord{\mathcal{O}}$ to represent the normal ordering of operator $\mathcal{O}$.
\begin{equation}\label{ope-cdc}
\begin{split}
&c^\dag(x)c(x')=\normord{e^{-i\phi(x)}}\normord{e^{i\phi(x')}}
=e^{-i\varphi^\dag(x)}e^{-i\varphi(x)} e^{i\varphi^\dag(x')}e^{i\varphi(x')}\\
=&\frac{i}{2\pi(x-x'+i0^+)}+\frac{\partial_x \phi}{2\pi}-\frac{i(x-x')}{4\pi} \Big[\normord{(\partial_x\phi)^2} +i\partial_x^2\phi
\Big]+\mathcal{O}\Big((x-x')^2\Big)\ ,
\end{split}
\end{equation} where the first term is the vacuum term. Taking the limit $x-x'\rightarrow 0$, we have
\beq
\normord{c^\dag(x)c(x)}=\frac{\partial_x \phi(x)}{2\pi}\ .
\eeq
For the chiral fermion kinetic term, we have 
\begin{equation}\label{ope-k}
\begin{split}
-ic^\dag(x)\partial_xc(x)&\approx i\left(\frac{c^\dag(x)+c^\dag(x')}{2}\right) \left(\frac{c(x)-c(x')}{x-x'}\right)=-\frac{i}{2}\left[\frac{c^\dag(x)c(x)-c^\dag(x')c(x')}{x-x'} +\frac{c^\dag(x')c(x)-c^\dag(x)c(x')}{x-x'}\right]\\
&=-\frac{1}{2\pi(x-x')^2}+\frac{1}{4\pi}\normord{(\partial_x\phi)^2},
\end{split}
\end{equation} which gives the bosonization mapping
\beq
-ic^\dagger(x)\partial_x c(x)=\frac{1}{4\pi}\normord{(\partial_x\phi)^2}
\eeq
in the limit $x\rightarrow x'$\ . 

Similarly, we can consider the OPE of charge $2e$ pairing term
\beq
\begin{split}
&c^\dag(x) c^\dag(x')=e^{-i\varphi^\dag(x)}e^{-i\varphi(x)} e^{-i\varphi^\dag(x')}e^{-i\varphi(x')}\\
=&e^{-i\varphi^\dag(x)}e^{-[\varphi(x),\varphi^\dag(x')]}e^{-i\varphi^\dag(x')} e^{-i\varphi(x)} e^{-i\varphi(x')} \approx -2\pi i (x-x'+i0^+)  \normord{e^{-2i\phi(x)}}\ .
\end{split}
\eeq
Taking the limit $x\rightarrow x'$, we thus have
\beq
\begin{split}
-ic^\dag(x)\partial_xc^\dagger(x)\approx i\left(\frac{c^\dag(x)+c^\dag(x')}{2}\right) \left(\frac{c^\dagger(x)-c^\dagger(x')}{x-x'}\right)=-\frac{i}{2}\left[\frac{c^\dag(x')c^\dag(x)-c^\dag(x)c^\dag(x')}{x-x'}\right]\approx 2\pi\normord{e^{-2i\phi(x)}}\ ,
\end{split}
\eeq and 
\beq
\begin{split}
-ic(x) \partial_x c(x) \approx i\left(\frac{c(x)+c(x')}{2}\right) \left(\frac{c(x)-c(x')}{x-x'}\right)=-\frac{i}{2}\left[\frac{c(x')c(x)-c(x)c(x')}{x-x'}\right] \approx 2\pi\normord{e^{2i\phi(x)}}\ .
\end{split}
\eeq 
\section{Exact solutions for chiral SY model with uniform couplings}\label{C}
In this appendix, we carefully derive exact solutions of chiral SY models 
\beq
\mathcal{L}_{SY}=-\frac{1}{4\pi} \sum_{i=1}^N\sum_{\mu,\nu=1}^{M-1}K_{\text{SU}(M)_1}^{\mu\nu}\partial_x \phi_{i\nu}( \partial_t +\partial_x) \phi_{i\mu}-\sum_{i\neq j}^N J_{ij}\bm{\mathcal{S}}_i(t,x) \cdot \bm{\mathcal{S}}_j(t,x)
\eeq
with uniform couplings $J_{ij}=J$ for $N=M=2$ and generic $N,M$.   
\subsection{$N=2, M=2$}\label{app:C1}
Starting with the total Lagrangian density
\beq
\mathcal{L}_{\text{dup}}=\mathcal{L}(\phi_1,\phi_2)+\mathcal{L}(\phi'_1,\phi'_2)\ ,
\eeq where each $\mathcal{L}$ is defined in Eq. (\ref{eq:LN=2M=2}). Defining $c_{j\pm}=e^{i\phi_{j}\pm i\phi'_{j}}$, the Lagrangian density becomes
\beq
\begin{aligned}
&\mathcal{L}_{\text{dup}}=i\displaystyle\sum_{j=1,2,\eta=\pm} c^\dagger_{j\eta} (\partial_t+\partial_x) c_{j\eta}+\mathcal{L}_{\text{int}}\ , \\
&\mathcal{L}_{\text{int}}=-\frac{J_\perp}{2} (e^{2i(\phi_1-\phi_2)}+e^{2i(\phi'_1-\phi'_2)})-\frac{J_z}{4\pi^2}(\partial_x \phi_1\partial_x \phi_2+\partial_x \phi'_1\partial_x \phi'_2)\\
&=-\frac{J_\perp}{2} (c_{1+}c_{1-}c_{2+}^\dagger c_{2-}^\dagger+c_{1+}c_{1-}^\dagger c_{2+}^\dagger c_{2-}+h.c.)-\frac{J_z}{2}(c^\dagger_{1+}c_{1+}c^\dagger_{2+}c_{2+}+c^\dagger_{1-}c_{1-}c^\dagger_{2-}c_{2-})\ .
\end{aligned}
\eeq

To further reveal the free nature of the present model, let us perform a rotation in the space of fermion operators, 
\begin{equation}
    \begin{aligned}
d_{\pm,\rho}=\frac{c_{1\pm}+ic_{2\pm}}{\sqrt{2}}=e^{i\theta_{\pm,\rho}}\ ,\\
d_{\pm,\sigma}=\frac{c_{1\pm}-ic_{2\pm}}{\sqrt{2}}=e^{i\theta_{\pm,\sigma}}\ .
\end{aligned}
\end{equation}

In this basis,
\beq
\begin{split}
&c_{1\pm}^\dagger c_{1\pm}c_{2\pm}^\dagger c_{2\pm}=\frac{(d_{\pm,\rho}^\dagger+d_{\pm,\sigma}^\dagger)(d_{\pm,\rho}+d_{\pm,\sigma})}{2}\frac{(d_{\pm,\rho}^\dagger-d_{\pm,\sigma}^\dagger)(d_{\pm,\rho}-d_{\pm,\sigma})}{2}=d_{\pm,\rho}^\dagger d_{\pm,\rho}d_{\pm,\sigma}^\dagger d_{\pm,\sigma}\ ,\\
&c_{1+}c_{1-}c_{2+}^\dagger c_{2-}^\dagger+c_{1+}c_{1-}^\dagger c_{2+}^\dagger c_{2-}+H.c.=(d_{+,\rho}^\dagger d_{+,\rho}-d_{+,\sigma}^\dagger d_{+,\sigma})(d_{-,\rho}^\dagger d_{-,\rho}-d_{-,\sigma}^\dagger d_{-,\sigma})
\end{split}
\eeq
and our Lagrangian density becomes
\beq
\begin{split}
&\mathcal{L}_{\text{dup}}=i\displaystyle\sum_{\eta=\pm,\alpha=\rho,\sigma} d^\dagger_{j,s} (\partial_t+\partial_x) d_{j,s}+\mathcal{L}_{int}\\
&=-\frac{1}{4\pi}\displaystyle\sum_{\eta=\pm,\alpha=\rho,\sigma}\partial_x \theta_{j,s} (\partial_t+\partial_x) \theta_{j,s}-\frac{J_z}{8\pi^2}[\partial_x \theta_{+,\rho} \partial_x \theta_{+,\sigma}+\partial_x \theta_{-,\rho} \partial_x \theta_{-,\sigma}]-\frac{J_\perp}{8\pi^2}(\partial_x \theta_{+,\rho}-\partial_x \theta_{+,\sigma})(\partial_x \theta_{-,\rho}-\partial_x \theta_{-,\sigma})\ .
\end{split}
\eeq 
This converts the model into a free boson model. Writing in $\bm{\theta}=(\theta_{+,\rho},\theta_{+,\sigma}, \theta_{-,\rho},\theta_{-,\sigma})^T$ basis, the velocity matrix for $\theta$ fields is
\beq
V=\begin{pmatrix}
1 & \frac{J_z}{4\pi} & \frac{J_{\perp}}{4\pi} & -\frac{J_{\perp}}{4\pi} \\
\frac{J_z}{4\pi} & 1 &  -\frac{J_{\perp}}{4\pi} &  \frac{J_{\perp}}{4\pi}\\
\frac{J_{\perp}}{4\pi} & -\frac{J_{\perp}}{4\pi}    & 1 & \frac{J_z}{4\pi}\\
-\frac{J_{\perp}}{4\pi} & \frac{J_{\perp}}{4\pi}    & \frac{J_z}{4\pi} & 1
\end{pmatrix}\ ,
\eeq
and the Lagrangian takes the form $\mathcal{L}_{\text{dup}}=-\frac{1}{4\pi}\left(\partial_t\bm{\theta}^T\partial_x\bm{\theta}+\partial_x\bm{\theta}^TV\partial_x\bm{\theta}\right)$.  By changing to the following basis

\begin{equation}\label{app:triality}
\widetilde{\theta}_{\eta\eta'}=(\theta_{+,\rho}+\eta\theta_{+,\sigma}+\eta'\theta_{-,\rho}+\eta\eta'\theta_{-,\sigma})/2\ ,
\end{equation} 
such that the velocity matrix $V$ is diagonalized, we finally get a free boson theory with velocities 
\begin{equation}
(u_{-+},u_{--},u_{++},u_{+-})=\left(1+\frac{J_\perp}{2\pi}-\frac{J_z}{4\pi},1-\frac{J_\perp}{2\pi}-\frac{J_z}{4\pi}, 1+\frac{J_z}{4\pi}, 1+\frac{J_z}{4\pi}\right)\ ,
\end{equation}
and the action
\beq\label{seq:Ldup-diag}
\mathcal{L}_{\text{dup}}=-\frac{1}{4\pi} \sum_{\eta,\eta'=\pm} \partial_x \widetilde{\theta}_{\eta\eta'} (\partial_t+u_{\eta\eta'} \partial_x) \widetilde{\theta}_{\eta\eta'}\ .
\eeq

\subsection{General $N,M$}\label{C2}

For the derivations here, it is more convenient to work with Hamiltonians than Lagrangians. The Hamiltonian density for $N$ copies of decoupled SU$(M)_1$ WZW models with ``speed of light" $1$ is
\beq
\mathcal{H}^{\otimes^N SU(M)_1}_{V=1}=\frac{1}{4\pi} \sum_{i=1}^N\sum_{\mu,\nu=1}^{M-1}K_{\text{SU}(M)_1}^{\mu\nu}\partial_x \phi_{i\nu}\partial_x \phi_{i\mu}\ ,
\eeq 
where the subindex $V=1$ denotes that the ``speed of light" is $1$. The Hamiltonian density can be equivalently expressed in the Sugawara form \cite{DiFrancesco1997} in terms of products of $SU(M)_1$ current densities as
\beq
\mathcal{H}^{\otimes^N SU(M)_1}_{V=1}= \frac{2\pi}{M+1}\sum_{i=1}^N\bm{\mathcal{S}}_i \cdot \bm{\mathcal{S}}_i\ .
\eeq
By noting that $\bm{\mathcal{S}}_{\text{tot}}=\sum_{i=1}^N\bm{\mathcal{S}}_{i}$ generates an $\mathfrak{su}(M)_N$ Kac-Moody algebra, and following the coset decomposition in Eq.~(\ref{coset}), we can rewrite $\mathcal{H}^{\otimes^N SU(M)_1}_{V=1}$ as
\beq \label{decom}
\mathcal{H}^{\otimes^N SU(M)_1}_{V=1}=\mathcal{H}^{\mathbb{G}_{M,N}}_{V=1}+\mathcal{H}^{SU(M)_N}_{V=1}\ .
\eeq 
According to the Sugawara construction \cite{DiFrancesco1997}, the Hamiltonian density $\mathcal{H}^{SU(M)_N}_{V=1}$ of the SU$(M)_N$ WZW model with ``speed of light" can be expressed by its current density $\bm{\mathcal{S}}_{\text{tot}}$ as
\beq
\mathcal{H}^{SU(M)_N}_{V=1}=\frac{2\pi}{M+N}\bm{\mathcal{S}}_{\text{tot}} \cdot \bm{\mathcal{S}}_{\text{tot}}\ .
\eeq 

We now examine the full interacting Hamiltonian density $\mathcal{H}$ of our model, which contains an interaction term $J\sum_{i\neq j}\bm{\mathcal{S}}_i \cdot \bm{\mathcal{S}}_j$ in addition to $\mathcal{H}^{\otimes^N SU(M)_1}_{V=1}$ in Eq. (\ref{decom}). It can be rewritten as
\begin{equation}
\begin{split}
&\mathcal{H}=\mathcal{H}^{\otimes^N SU(M)_1}_{V=1}+J\sum_{i\neq j}\bm{\mathcal{S}}_i \cdot \bm{\mathcal{S}}_j =\frac{2\pi}{M+1}\sum_{i=1}^N\bm{\mathcal{S}}_i \cdot \bm{\mathcal{S}}_i-J\sum_{i=1}^N\bm{\mathcal{S}}_i \cdot \bm{\mathcal{S}}_i+J\bm{\mathcal{S}}_{\text{tot}} \cdot \bm{\mathcal{S}}_{\text{tot}} \\
&=\left(\frac{2\pi}{M+1}-J\right)\sum_{i=1}^N\bm{\mathcal{S}}_i \cdot \bm{\mathcal{S}}_i+J\bm{\mathcal{S}}_{\text{tot}} \cdot \bm{\mathcal{S}}_{\text{tot}} = \left( 1-\frac{(M+1)J}{2\pi} \right)\mathcal{H}^{\otimes^N SU(M)_1}_{V=1}+\frac{(M+N)J}{2\pi} \mathcal{H}^{SU(M)_N}_{V=1} \\
&=\left( 1-\frac{(M+1)J}{2\pi} \right)\left(\mathcal{H}^{\mathbb{G}_{M,N}}_{V=1}+\mathcal{H}^{SU(M)_N}_{V=1}\right)+\frac{(M+N)J}{2\pi} \mathcal{H}^{SU(M)_N}_{V=1} \\
&=v_{\mathbb{G}_{M,N}} \mathcal{H}^{\mathbb{G}_{M,N}}_{V=1} + v_{SU(M)_N} \mathcal{H}^{SU(M)_N}_{V=1}\ ,
\end{split}
\end{equation}
where the velocities are the effective ``speed of light" given by
\beq
v_{\mathbb{G}_{M,N}}=v_J=1-\frac{(M+1)J}{2\pi}\ ,\qquad v_{SU(M)_N}=1+\frac{(N-1)J}{2\pi}\ ,
\eeq
with $v_J=1-\frac{(M+1)J}{2\pi}$ defined in the main text. Therefore, the chiral interaction yields two different ``speed of lights" for the $\mathbb{G}_{M,N}$ coset part and the $SU(M)_N$ WZW part respectively.

\section{Orthogonal transformation and ancillary fields}\label{D}
In this appendix, by adding ancillary fields, we show an explicitly transformation from $N$ copies of $M$ chiral fermion basis into another basis consisting of a $U(1)$ sector and a $SU(M)_1$ sector with corresponding boson fields. 

Starting with $N$ copies of $M$ chiral fermions, bosonization gives us $f_{i,\mu}|_{i=1,\mu=1}^{N,M}=e^{i\widetilde{\phi}_{i,\mu}}$, the following explicit orthogonal transformation decomposes each copy into a $U(1)$ sector and a $SU(M)_1$ sector:
\begin{equation}
\begin{pmatrix}
\phi_{i,1}\\
\vdots\\
\phi_{i,M-1}\\
\phi'_{i,\rho}
\end{pmatrix}\\
=\textbf{O} \begin{pmatrix}
\widetilde{\phi}_{i,1}\\
\vdots\\
\widetilde{\phi}_{i,M-1}\\
\widetilde{\phi}_{i,M}
\end{pmatrix}
\end{equation} for all $1\leq i\leq N$ where 
\begin{equation}
\textbf{O}=\begin{pmatrix}
\frac{1}{\sqrt{2}} & -\frac{1}{\sqrt{2}} & 0 & 0 & \ldots & 0 \\
\frac{1}{\sqrt{6}} & \frac{1}{\sqrt{6}} & -\frac{2}{\sqrt{6}}  & 0 &\ldots & 0 \\
\vdots & \vdots  & & \ddots\\
\frac{1}{\sqrt{(M-1)M}} & \frac{1}{\sqrt{(M-1)M}} & & \ldots & & -\frac{M-1}{\sqrt{(M-1)M}}\\
\frac{1}{\sqrt{M}} & \frac{1}{\sqrt{M}} & & \ldots & & \frac{1}{\sqrt{M}} 
\end{pmatrix}
\end{equation}
or more concisely,
\begin{equation}
\begin{split}
&\phi_{i,\mu}|_{1\leq\mu \leq M-1}=\frac{1}{\sqrt{\mu(\mu+1)}}(\sum_{\nu=1}^\mu \widetilde{\phi}_{i,\nu}-\mu \widetilde{\phi}_{i,\mu+1})\ ,\\
&\phi'_{i,\rho}=\frac{1}{\sqrt{M}}\sum_{\mu=1}^M \widetilde{\phi}_{i,\mu}\ .
\end{split}
\end{equation}
\section{The large $N,M$ limit of chiral SY model in the enlarged complex fermion representation}\label{E}
In the main text Eq. (\ref{eq:Ltot-ancilla}) we showed that the 1+1d chiral Sachdev-Ye model with ancillary fields added map into a complex fermion model in the large $N,M$ limit:
\beq
\begin{split}
\mathcal{L}_{f}=i\sum_{i=1}^N\sum_{\mu=1}^M f^\dagger_{i,\mu} (\partial_t+\partial_x)f_{i,\mu}-\sum_{i<j}^N \sum_{\mu<\nu}^M J_{ij} f_{i,\mu}^\dagger f_{i,\nu} f_{j,\nu}^\dagger f_{j,\mu}\ ,
\end{split}
\eeq where $J\geq 0$, $J_{ij}=J_{ji}$ and 
 \beq
 \la J_{ij} \ra =0, \quad \la J_{ij}J_{i'j'}\ra=\frac{J^2}{(N-1)(M-1)} \delta_{ii'} \delta_{jj'}\ .
 \eeq
In this section, we calculate the two-point function and the OTOC of this complex fermion model, which gives the factor $\delta\mathcal{F}$ of the OTOC of anyons in Eq. (\ref{eq:otoc-expansion}).

\subsection{Two-point function}
\begin{figure}[h!]
    \centering
    \includegraphics[width=6.8in]{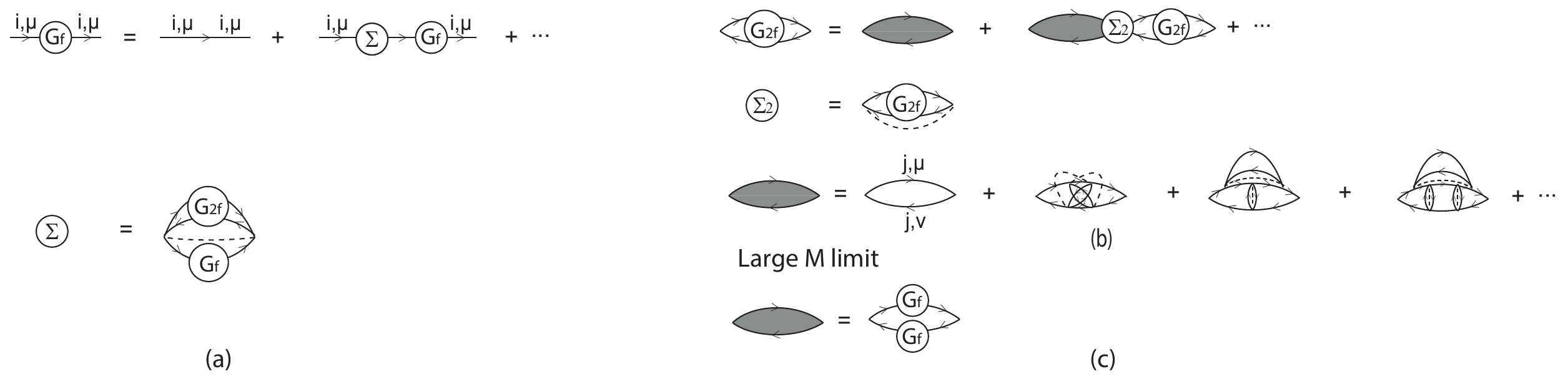}
    \caption{(a) Single fermion propagators $G_f$ (Eq.~(\ref{seq:G_f})) in the large $N$ limit. (b) Double fermion propagators $G_{2f}^{\mu\nu}$ (Eq.~\ref{seq:g2}) in the large $N$ limit. (c) In the large $M$ limit, the double fermion diagrams further simplify $G_{2f}=(G_f)^2$.  Dashed line denotes average over $J_{ij}$.
    }
    \label{fig:diag}
\end{figure}

We define the fermion two-point function at zero temperature as
\be
G_f(t,x)=\frac{1}{N}\sum_{i=1}^N \la f_{i,1}(t,x) f_{i,1}^\dag(0,0)\ra\ .
\label{seq:G_f}
\ee
In the large $N$ limit, dominant contributions to the two-point function are from two components - the single and double fermion propagators. They are summarized in Fig.~\ref{fig:diag} (a) and (b) respectively. The double fermion propagators are defined as
\beq\label{seq:g2}
G_{2f}^{\mu\nu}(t,x)=\frac{1}{N}\sum_{j=1}^N\la f_{j,\nu}(t,x) f_{j,\mu}^\dag(t,x) f_{j,\mu}(0,0)f_{j,\nu}^\dag(0,0) \ra
\eeq
Different from the usual chiral SYK physics\cite{Lian2019}, there are diagrams of orders of $1/M$ in chiral SY model in the large $N$ limit, for example the last three diagrams in Fig.~\ref{fig:diag} (b). In the large $M$ limit, to the leading order in $\frac{1}{M}$, $G_{2f}^{\mu\nu}(t,x)$ becomes independent on $\mu,\nu$ (thus the $\mu,\nu$ index is dropped in Fig. \ref{fig:diag}), and is given by $G_{2f}^{\mu\nu}(t,x)\equiv G_{2f}(t,x)=-G_f(t,x)G_f(-t,-x)=[G_f(t,x)]^2$ ($G_f(t,x)$ is an odd function in $t$ and $x$, due to the anticommuting fermion statistics). For diagrams in Fig.~\ref{fig:diag} (b), diagrams on both the first and third line will be dominated by the first diagram after the equality sign.  After that, it is clear that we recover the same leading Feynman diagrams in the large $N, M$ limit as that of the SYK models. This directly leads to the following Schwinger-Dyson (SD) equation for the two point function $G$ and the self energy $\Sigma$ at zero temperature:
\beq\label{seq:SD}
\frac{1}{G_f(\omega,k)}=-\omega+k-\Sigma(\omega,k),\quad \Sigma(t,x)=-J^2 G_{2f}(t,x) G_f(-t,-x)=J^2 [G_{f}(t,x)]^3. 
\eeq 
It is clear that $G(\omega,k)$ and $\Sigma(\omega,k)$ have scaling dimension $-1$ and $+1$, respectively. Scaling invariance at zero temperature thus allows us to constrain $G(\omega,k)$ and $\Sigma(\omega,k)$ into a scaling invariant form. By noting that the SD Eq. (\ref{seq:SD}) here has the same form as that of the chiral SYK model in Ref. \cite{Lian2019}, we follow the derivations from \cite{Lian2019}, and finally arrive at the real-space two point Green's function the large $N, M$ limit as
\be
G_f(t,x)=\frac{1}{2\pi i}\frac{1}{\sqrt{(u_+t-x-i0^+)(u_-t-x-i0^+)}}\ ,
\ee and the real space self-energy
\be
\Sigma(t,x)=-\frac{J^2}{8\pi^3} \frac{1}{[(u_+t-x-i0^+)(u_-t-x-i0^+)]^{3/2}}\ ,
\ee where the two velocities are $u_{\pm}=1\pm\frac{J}{2\pi}$.

\subsection{The OTOC and chaos in the large $N,M$ limit}\label{F2}

\begin{figure}[h!]
    \centering
    \includegraphics[width=5.5in]{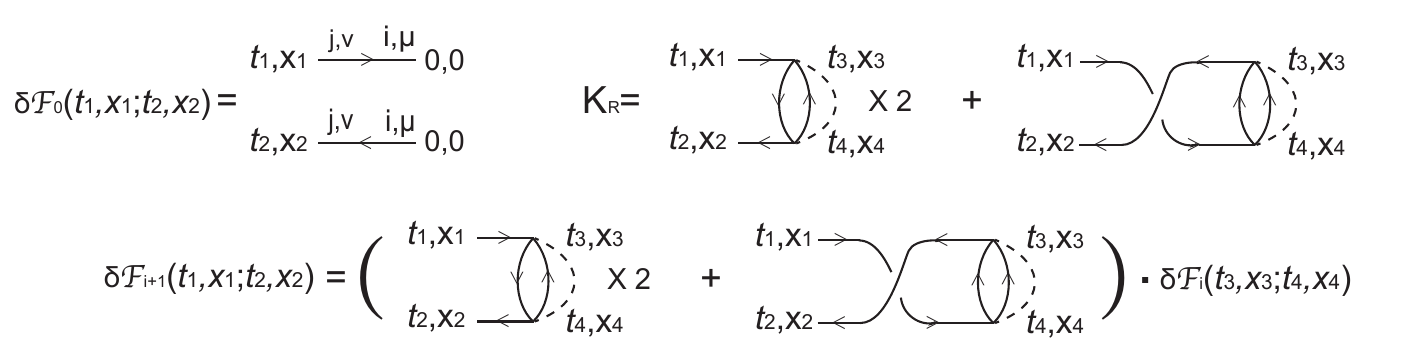}
    \caption{Ladder diagrams contributing $\delta \mathcal{F}$ of order $\frac{1}{NM}$. The dashed lines stand for averaging over $J_{ij}$. $\delta \mathcal{F}$  is approximately an eigenfunction of kernel $K_R$ with eigenvalue $1$ (Eq. (\ref{fig:S2}).)
    }
    \label{fig:diag2}
\end{figure}
In this subsection, following a similar derivation in Ref. \cite{Lian2019}, we calculate the regularized out-of-time-order four-point function in real time
\beq
\frac{1}{N^2M^2}\sum_{i,j=1}^N\sum_{\mu,\nu=1}^M \text{Tr}[yf_{j,\nu}^\dagger(t_1,x_1)yf_{i,\mu}(0,0)yf_{j,\nu}(t_2,x_2)yf_{i,\mu}^\dagger(0,0)]\ ,
\eeq where $y=e^{-\beta H/4}=\rho(\beta)^{1/4}$ separates the four fermion fields by a quarter of the thermal circle. The leading order contributions coming from Wick contracting the two $f_{j,\nu}$ and two $f_{i,\mu}$ are 
\be
-G_f^\beta(-i\frac{\beta}{2},0)G_f^\beta(t_2-t_1-i\frac{\beta}{2},x_2-x_1)\ .
\ee
The next order contributions come from contracting $f_{j,\nu}$ with $f_{i,\mu}$
\be
-\frac{1}{NM}\delta\mathcal{F}(t_1,x_1;t_2,x_2)\ ,
\ee
where $\delta\mathcal{F}$ is the function appearing in Eq. (\ref{eq:otoc-expansion}). Writing $\delta\mathcal{F}(t_1,x_1;t_2,x_2)=\sum_{i=0}^{\infty} \delta\mathcal{F}_i(t_1,x_1;t_2,x_2)$ order by order in diagrams, the first diagram is given in Fig.~\ref{fig:diag2}, where
\be
\delta\mathcal{F}_0(t_1,x_1;t_2,x_2)=G_{fR}^\beta(t_1,x_1)G_{fR}^\beta(t_2,x_2)
\ee and the kernel acts as 
\be\label{seq:KR1}
K_R \circ \delta\mathcal{F}_i=\delta\mathcal{F}_{i+1}
\ee such that 
\be\label{seq:KR2}
\delta\mathcal{F}=\frac{1}{1-K_R}\delta\mathcal{F}_0\ ,
\ee
where the retarded kernel $K_R$ is
\beq\label{seq:KR}
K_R(t_1,x_1;\cdots;t_4,x_4)=2J^2 G_{fR}^\beta(t_{13},x_{13})G_{fR}^\beta(t_{24},x_{24})(G_{fW}^\beta)^2(t_{34},x_{34})+J^2G_{fR}^\beta(t_{14},x_{14})G_{fR}^\beta(t_{23},x_{23})(G_{fW}^\beta)^2(t_{34},x_{34}). 
\eeq For simplicity, we have used the fact that $G^\beta_{fA}(t,x)=G^\beta_{fR}(-t,-x)$ for propagators with $(t_{31},x_{31})$ and $(t_{41},x_{41})$. Here we defined $t_{ij}=t_i-t_j$ and $x_{ij}=x_i-x_j$. $G_{fR}^\beta$ ($G^\beta_{fA}$) is the retarded (advanced) Greens function and $G_{fW}^\beta$ is the Wightman correlator with half thermal circle separation.
They are explicitly
\beq
G_{fR}^\beta(t,x)=\frac{1}{\beta\sqrt{u_+ u_-}}\frac{\Theta(t-u^{-1}_+x)\Theta(u^{-1}_-x-t)}{\sqrt{\sinh[\frac{\pi}{\beta}(t-u^{-1}_+x)]\sinh[\frac{\pi}{\beta}(u^{-1}_-x-t)]}}
\eeq and
\beq
G_{fW}^\beta(t,x)=\frac{1}{2\beta\sqrt{u_+ u_-}}\frac{1}{\sqrt{\cosh[\frac{\pi}{\beta}(t-u^{-1}_+x)]\cosh[\frac{\pi}{\beta}(u^{-1}_-x-t)]}}\ .
\eeq 

At large time $\beta < t < \beta \log NM$ within the Lyapunov regime, we expect $\delta\mathcal{F}$ to grow exponentially and be dominated by $\delta\mathcal{F}_j$ with large $j$. Eqs. (\ref{seq:KR1}), (\ref{seq:KR2}) thus indicate that $\delta\mathcal{F}$ satisfies the following self-consistent equation
\beq
\delta\mathcal{F}(t_1,x_1;t_2,x_2)=\int dt_3 dx_3 dt_4 dx_4 K_R(t_1,x_1;\cdots;t_4,x_4)\delta\mathcal{F}(t_3,x_3;t_4,x_4)\ ,
\label{fig:S2}
\eeq  
This shows that $\delta \mathcal{F}$ is an eigenfunction of the kernel $K_R$. 
 
The expression of the kernel $K_R$ here is the same as that in Ref. \cite{Lian2019}, where all the eigenfunctions of the kernel $K_R$ have been worked out. Due to translational symmetry, these eigenfunctions are plane waves. When $t_1=t_2=t,x_1=x_2=x$, it assumes the form $e^{\frac{2\pi}{\beta}[\varkappa(p)t+ipx]}$ with 
\beq
\varkappa(p)=\frac{-\mathcal{J}-ip(1-\mathcal{J}^2)+\mathcal{J}\sqrt{3(1-\mathcal{J}^2)+(\mathcal{J}+(1-\mathcal{J}^2)ip)^2}}{1-\mathcal{J}^2}, \quad \mathcal{J}=\frac{J}{2\pi}\ .
\eeq
The reduced OTOC function $\delta\mathcal{F}$ is a sum of $e^{\frac{2\pi}{\beta}[\varkappa(p)t+ipx]}$ with weight $\sim \frac{1}{\cos[\pi \varkappa(p)/2]}$\cite{Gu2019} such that
\beq\label{seq:dF}
\delta\mathcal{F}(t,x) \sim \int_{-\infty}^{\infty}dp \frac{e^{\frac{2\pi}{\beta}[\varkappa(p)t+ipx]}}{\cos[\pi \varkappa(p)/2]}\ .
\eeq
Along a fixed velocity $x=vt$, depending on the velocity $v$, the above integral is either dominated by a saddle point or a pole of the function $\varkappa(p)$ in the complex plane of $p$. We refer the readers to Ref. \cite{Lian2019} Sec. 4.3 for the detailed calculation of Eq. (\ref{seq:dF}), while here we only summarize the results.

For a fixed velocity $v$, if $p_v$ is the dominant saddle point or pole of function $\varkappa(p)$, the function $\delta \mathcal{F}$ will grow as
\beq
\delta\mathcal{F}(t,x=vt)\sim e^{\lambda_v t}\ ,
\eeq
where the velocity dependent Lyapunov exponent (VDLE) is given by
\begin{equation}
\lambda_v=\frac{2\pi}{\beta}[\varkappa(p_v)+ip_v v]\ .
\end{equation}
There are three regimes separated by two velocities $v_*=\frac{2-2\mathcal{J}^2}{2-\mathcal{J}}$, $v_*'=\frac{2-2\mathcal{J}^2}{2+\mathcal{J}}$:

(i) When $v_*'<v<v_*$, the integral (\ref{seq:dF}) is dominated by a saddle point satisfying $\varkappa'(p_v)+iv=0$. Such a saddle-dominated regime is known as the diffusive regime \cite{mezei2019}, and the corresponding $\lambda_v$ is given in the main text Eq. (\ref{eq:Lv}).

(ii) When $u_-<v<v_*'$ (or $v_*<v<u_+$), the integral (\ref{seq:dF}) is dominated by a pole at $p_v=-\frac{i}{u_-}$ (or $p_v=\frac{i}{u_+}$). Such a pole-dominated regime is known as the ballistic regime \cite{mezei2019}, and the corresponding $\lambda_v$ expressions in the main text Eq. (\ref{eq:Lv}).

\twocolumngrid
\bibliographystyle{unsrt}

\end{document}